%% file: main.tex
\ifx\destination\undefined
  \def\destination{arxiv}  % publisher,arxiv
\fi

\RequirePackage{snapshot}
\PassOptionsToPackage{dvipsnames}{xcolor}
\PassOptionsToPackage{colorlinks=True,allcolors=RoyalBlue}{hyperlink}

\def\arxiv{arxiv}
\def\publisher{publisher}

\ifx\destination\arxiv
  \documentclass{article}
  \pdfoutput=1  % ensure pdflatex on arXiv
  \usepackage{aux/preprint}
  \usepackage{natbib}
  \usepackage{authblk}
  \let\address\affil
\else
  \documentclass[preprint,3p,authoryear,lefttitle]{elsarticle}
  \journal{Icarus}
\fi

\usepackage{amsmath}
\usepackage{aux/aas_macros}
\usepackage{graphicx}
\usepackage[title]{appendix}
\usepackage{booktabs}
\usepackage[skip=4pt]{caption}
\usepackage{here}
\usepackage[utf8]{inputenc}
\usepackage{pgf}
\usepackage{tikz}
\usepackage{url}
\usepackage{xspace}
\usepackage[colorlinks=True, allcolors=RoyalBlue]{hyperref}
\usepackage[nameinlink]{cleveref}

\input{aux/macros}
\input{aux/stats}

\begin{document}

% ------
\title{Asteroid phase curves from ATLAS dual-band photometry
  \ifx\destination\publisher
      \footnote{The catalogue of absolute magnitudes and phase parameters is publicly
      available at the CDS via anonymous ftp to \url{cdsarc.u-strasbg.fr}
      (\url{130.79.128.5}) or via \url{http://cdsarc.u-strasbg.fr/viz-bin/qcat?VII/288}
    }
  \fi
}

% Authors
\author[1,2]{Max Mahlke}
\ifx\destination\publisher
  \corref{cor}
  \cortext[cor]{Corresponding author}
  \ead{max.mahlke@oca.eu}  % email address of *most recent* author
\fi

\author[1]{Benoit Carry}
\author[3]{Larry Denneau}

% Affiliations
\address[1]{Universit{\'e} C{\^o}te d'Azur, Observatoire de la C{\^o}te d'Azur,
              CNRS, Laboratoire Lagrange, France}
\address[2]{CAB (INTA-CSIC), Campus ESAC (ESA), Villanueva de la Ca\~nada,
              Madrid, Spain}
\address[3]{University of Hawaii, 2680 Woodlawn Dr., Honolulu HI 96822 USA}

% Abstract
\ifx\destination\arxiv
  \twocolumn[
    \begin{@twocolumnfalse}
      \maketitle
    \begin{abstract}
      \input{frontback/abstract}
    \end{abstract}
  \end{@twocolumnfalse}
  ]
\else
  \begin{frontmatter}
    \begin{abstract}
      \input{frontback/abstract}
    \end{abstract}
  \end{frontmatter}
\fi

% ------
% Introction
\input{sections/section1}
% Methodology
\input{sections/section2}
% Results
\input{sections/section3}
% Taxonomy
\input{sections/section4}

% Asteroid Families
\input{sections/section5}

% Error sources
\input{sections/section6}
% Conclusion
\input{sections/section7}
% ------

\input{frontback/ack}

%------
% Bibliography
\ifx\destination\arxiv
  \bibliographystyle{aux/arxiv.bst}
\else
  \bibliographystyle{aux/publisher.bst}
  \biboptions{authoryear}
\fi

\bibliography{aux/bib}

\clearpage
% ------
% Appendix
\input{frontback/appendix}
\end{document}

%% file: aux/macros.tex
\newcommand{\nuna}[2]{(#1)\,\textit{#2}} % For asteroid number-name pairs

\newcommand{\cyan}{\textit{cyan}\xspace}
\newcommand{\orange}{\textit{orange}\xspace}

\newcommand{\GG}{$G_1,\,G_2$\xspace}
\newcommand{\GS}{$G_{12}^*$\xspace}
\newcommand{\HGG}{$H,\,G_1,\,G_2$\xspace}
\newcommand{\HGNS}{$H,\,G_{12}$\xspace}
\newcommand{\HGS}{$H,\,G_{12}^*$\xspace}

\usepackage{catoptions}
\makeatletter

\def\Autoref#1{%
  \begingroup
  \edef\reserved@a{\cpttrimspaces{#1}}%
  \ifcsndefTF{r@#1}{%
    \xaftercsname{\expandafter\testreftype\@fourthoffive}
      {r@\reserved@a}.\\{#1}%
  }{%
    \ref{#1}%
  }%
  \endgroup
}
\def\testreftype#1.#2\\#3{%
  \ifcsndefTF{#1autorefname}{%
    \def\reserved@a##1##2\@nil{%
      \uppercase{\def\ref@name{##1}}%
      \csn@edef{#1autorefname}{\ref@name##2}%
      \autoref{#3}%
    }%
    \reserved@a#1\@nil
  }{%
    \autoref{#3}%
  }%
}
\makeatother

\newcommand\inputpgf[2]{{
\let\includegraphicsWithoutPath\includegraphics
\renewcommand{\includegraphics}[2][]{\includegraphicsWithoutPath[##1]{#1/##2}}
\input{#1/#2}
}}

%% file: aux/stats.tex
\newcommand\phminlimit{3\xspace}
\newcommand\Nlimit{50\xspace}

\newcommand\numbphasecurves{127,012\xspace}
\newcommand\numbphasecurvestaxlimit{61,184\xspace}
\newcommand\numbuniqueasteroidswithphasecurve{94,777\xspace}
\newcommand\numbphasecurvesc{36,441\xspace}
\newcommand\numbphasecurveso{90,571\xspace}
\newcommand\numbfailedfits{43\xspace}

\newcommand\asteroidswithclasstaxlimit{19,708\xspace}

\newcommand\asteroidswithalbedotaxlimit{14,384\xspace}
\newcommand\gscbelow{42}
\newcommand\gsobelow{50}
\newcommand\numbproperelements{93,200\xspace}
\newcommand\numbatlasasteroids{180,025\xspace}
\newcommand\numbatlasasteroidsc{179,719\xspace}
\newcommand\numbatlasasteroidso{180,025\xspace}
\newcommand\numbtwodeg{124,072\xspace}

\newcommand\numbdamitasteroids{2,407\xspace}
\newcommand\numbdamitatlas{720\xspace}
\newcommand\numbdamitatlasphasecurves{917\xspace}

\newcommand\percbelowtenmag{0.8}
\newcommand\fammemberlimit{500\xspace}

\newcommand\numbfamiliesoverlimit{10\xspace}
\newcommand\listoffamilies{~\nuna{4}{Vesta}, ~\nuna{5}{Astraea}, ~\nuna{10}{Hygiea}, ~\nuna{15}{Eunomia}, ~\nuna{24}{Themis}, ~~\nuna{93}{Minerva}, ~~\nuna{135}{Hertha}, ~~\nuna{158}{Koronis}, \nuna{170}{Maria}, and \nuna{221}{Eos}}
\newcommand\refVesta{10.1086115658}
\newcommand\refAstraea{10.1093mnrasstx843}
\newcommand\refHygiea{10.1093mnrasstt437}
\newcommand\refEunomia{10.1016j.icarus.2010.02.011}

\newcommand\refThemis{10.1016j.icarus.2004.10.002}
\newcommand\refKoronis{1984PhDT.........3T}

\newcommand\refMinerva{10.1016j.icarus.2004.10.002}
\newcommand\refHertha{10.1016j.icarus.2015.01.012}
\newcommand\refMaria{10.1006icar.1997.5749}
\newcommand\refEos{Masiero_2014}

\newcommand\taxVesta{V}
\newcommand\taxAstraea{S}
\newcommand\taxHygiea{C}
\newcommand\taxEunomia{S}

\newcommand\taxThemis{C}
\newcommand\taxMinerva{S}
\newcommand\taxHertha{S}
\newcommand\taxKoronis{S}
\newcommand\taxMaria{S}
\newcommand\taxEos{K}

\newcommand\meandiffug{0.03}
\newcommand\meandiffgr{-0.03}
\newcommand\meandiffri{-0.04}
\newcommand\meandiffiz{0.02}

%% file: frontback/abstract.tex
Asteroid phase curves are used to derive fundamental physical properties
  through the determination of the absolute magnitude $H$.
The upcoming visible
  \emph{Legacy Survey of Space and Time} (LSST) and
  mid-infrared
  \emph{Near-Earth Object Surveillance Mission} (NEOSM) surveys rely
  on these absolute magnitudes to derive the colours and albedos of millions of
  asteroids.
Furthermore, the shape of the phase curves reflects their surface compositions,
  allowing for conclusions on their taxonomy.
We derive asteroid phase curves from dual-band photometry acquired by the
  \emph{Asteroid Terrestrial-impact Last Alert System} telescopes.
Using Bayesian parameter inference, we retrieve the absolute magnitudes and
  slope parameters of
  \numbphasecurves phase curves of
  \numbuniqueasteroidswithphasecurve asteroids in the photometric \HGG- and
  \HGS-systems.
The taxonomic complexes of asteroids separate in the observed
  \GG-distributions, correlating with their mean visual albedo.
This allows for differentiating the X-complex into the P-, M-, and E-complexes
  using the slope parameters as alternative to albedo measurements.
Further, taxonomic misclassifications from spectrophotometric datasets as well
  as interlopers in dynamical families of asteroids reveal themselves in
  \GG-space.
The \HGS-model applied to the serendipitous observations is unable to resolve
  target taxonomy.
The \GG phase coefficients show wavelength-dependency for the majority of
  taxonomic complexes.
Their values allow for estimating the degree of phase reddening of the spectral
  slope.
The uncertainty of the phase coefficients and the derived absolute magnitude is
  dominated by the observational coverage of the opposition effect rather than
  the magnitude dispersion induced by the asteroids' irregular shapes and
  orientations.
Serendipitous asteroid observations allow for reliable phase curve
  determination for a large number of asteroids.
To ensure that the acquired absolute magnitudes are suited for colour
  computations, it is imperative that future surveys densely cover the opposition
  effects of the phase curves, minimizing the uncertainty on $H$.
The phase curve slope parameters offer an accessible dimension for taxonomic
  classification, correlating with the albedo and complimentary to the spectral
  dimension.

%% file: sections/section1.tex
\section{Introduction}

The absolute magnitude $H$ of asteroids is defined as their apparent Johnson
$V$-band magnitude observed at zero degree solar phase angle and reduced to
1\,AU distance from both the Sun and the Earth, averaged over a full period of
their rotation. The phase angle $\alpha$ is the angle between the Sun, the
asteroid, and the observer. The reduced magnitude $V(\alpha)$ is calculated
from the observed apparent magnitude $m$ as
\begin{equation}%
  \label{equ:reduced_apparent_mag}
  V(\alpha) = m + 5\log(r\Delta)\,,
\end{equation}
where $r$ is the distance between the asteroid and the Sun at the epoch of
observation and $\Delta$ the respective distance between the asteroid and Earth.
$V(\alpha)$ is referred to as the phase curve, and, by definition,
\mbox{$H=V(0)$}.

The inference of principal physical parameters of minor bodies requires accurate
knowledge of their absolute magnitudes. Their diameter $D$ and visual geometric
albedo $p_V$ are related to $H$ by \citep{2002aste.book..205H}
\begin{equation}
  \label{equ:hdp}
  \log_{\mathrm{10}}D = 3.1236 - 0.2H - 0.5\log_{\mathrm{10}}p_V\,.
\end{equation}
Any uncertainty in $H$ enters logarithmically in the derivation of the physical
properties. The diameters and visual albedos of more than 100,000 Main Belt
asteroids observed with NASA's \emph{Wide-field Infrared Survey Explorer} (WISE)
carry 20\% and 40\% accuracy, under the assumption that the referenced absolute
magnitudes are accurate \citep{2011ApJ...741...68M, 2015ApJ...814..117N}. 
NASA's planned \emph{~Near-~Earth ~Object ~Surveillance Mission}
\citep[NEOSM, previously NEOCam,][]{2019LPI....50.3175G} aims to extend this 
catalogue by an order of magnitude, thereby vastly increasing the demand for
accurate determinations of $H$.

Deriving $H$ in different wavelength bands further offers the consolidation of
asteroid photometry obtained at different epochs for colour computation and
subsequent taxonomic classification. This is vital for the upcoming \emph{Legacy
Survey of Space and Time} (LSST) executed at the Vera C. Rubin Observatory. The
LSST aims to provide catalogues of photometric variability and colours for
millions of minor bodies \citep{10.1007s11038-009-9305-z}. The latter
necessitates either quasi-simultaneous multi-band observations of a single
target or reduction of the observed magnitudes to zero phase angle
\citep{10.1111j.1365-2966.2004.07426.x,10.1016j.icarus.2009.02.003}. Given the
numerous competing science cases of the LSST and its planned operations with two
filters per night at most, the Solar System science community cannot rely on the
realization of the required observation cadence alone. Instead, $H$ must be
derived in each band by fitting the observed phase curves to obtain the colours. 

The definition of $H$ requires asteroid magnitudes at zero
degree phase angle. This is practically difficult to achieve, hence $H$ is
instead extrapolated from photometric observations acquired close to opposition,
but at non-zero phase angles, by means of phase curve modelling. We summarize
here the most basic modelling advances and refer to
\citet{2002MmSAI..73..716M} and \citet{2015aste.book..129L} for detailed
reviews.

In first order, an asteroid's apparent brightness increases linearly with
decreasing phase angle. The slope of the phase curve is dictated by mutual
shadowing of the surface particles, which in turn depends on the surface
properties like shape, roughness and porosity. When observing an asteroid close
to opposition, a nonlinear brightness surge occurs, referred to as
\emph{opposition effect} \citep{10.1086146166}. In 1985, the \emph{International
Astronomical Union} (IAU) adopted the $H,\,G$-magnitude system, where $G$
describes the overall slope of the phase curve \citep{1989aste.conf..524B}. The
$H$,\,$G$-system successfully describes phase curves in large ranges of the
phase space, however, it fails to reproduce the opposition effect, especially
for exceptionally dark or bright objects \citep{10.1006icar.2000.6410}. In 2010,
the $H,\,G$ system was extended by \citet{10.1016j.icarus.2010.04.003} to the
three-parameter \HGG-system, 
\begin{equation}%
  \label{equ:hg1g2}
  \begin{split}
    V(\alpha) = H - 2.5\log_{10}[G_1\Phi_1(\alpha) &+G_2\Phi_2(\alpha) \\
                                                   &+(1-G_1-G_2)\Phi_3(\alpha)]\,,
  \end{split}
\end{equation}
\noindent where the $\Phi_i$ are basis functions describing the linear part
(subscripts 1 and 2) and the opposition effect (subscript 3). For low-accuracy
and sparsely-sampled phase curves, the authors propose the \HGNS-system, later
refined by \citet{10.1016j.pss.2015.08.010} to the \HGS-system, where 
\begin{equation}%
  \label{equ:g12star}
  (G_1, G_2) =   \begin{pmatrix} 0 \\ 0.53513350 \end{pmatrix} + G_{12}^*
  \begin{pmatrix} 0.84293649 \\ -0.53513350 \end{pmatrix}\,.
\end{equation}
Taking into account the physical constraint that asteroids get fainter with
increasing phase angle, we confine the \GG-space using
\autoref{equ:hg1g2} to
\begin{subequations}
  \label{equ:constraints}
  \begin{align}
    G_1, G_2 &\geq 0  \label{equ:con_g1g2}\,,\\
    1-G_1-G_2&\geq 0 \label{equ:con_g1andg2}\,.
  \end{align}
\end{subequations}

We gain physical interpretability of the phase coefficients by expressing the
photometric slope $k$ between 0\,deg and 7.5\,deg following
\citet{10.1016j.icarus.2010.04.003} as 
\begin{equation}%
  \label{equ:k}
  k = -\frac{1}{5\pi} \frac{30\,G_1+9\,G_2}{G_1+G_2}\,,
\end{equation}
and the size of the opposition effect $\zeta - 1$ as 
\begin{equation}%
  \label{equ:z}
  \zeta - 1 = \frac{1-G_1-G_2}{G_1+G_2}\,,
\end{equation}
where $\zeta$ is the ratio of the amplitude of the opposition effect and the
background intensity. $k$ is in units of mag/rad, while $\zeta-1$ gives the
contribution of the opposition effect to the absolute magnitude in units of
mag.
\citet{10.1006icar.2000.6410} showed that the opposition effect and the
photometric slope correlate with the albedo. The former peaks for
moderate-albedo asteroids, while minor bodies with high- and low-albedo
asteroids display smaller opposition effects. $k$ is proportional to the albedo,
with dark asteroids exhibiting steeper phase curves than bright minor bodies.

The derivation of accurate phase curve parameters requires multiple
observational campaigns at different solar elongations of a single target. The
observations need to account for the modulation of the apparent magnitude by the
asteroid's irregular shape and rotation, in addition to possible offsets due to
varying aspect angles when combining data from distinct apparitions. Examples of
targeted campaigns can be found in \citet{10.1016s0032-06339700131-1,
2002Icar..155..365S,10.1016j.icarus.2008.03.015,10.1016j.pss.2015.11.007}. 
The number of asteroids with accurate and reduced phase curves available remains
in the lower hundreds due to the requirements of extensive telescope time and
asteroid shape models. 

To obtain catalogues of phase curve parameters in the order of magnitude
required for future large-scale surveys, serendipitous asteroid observations
need to be exploited. \citet{10.1016j.jqsrt.2011.03.003} determined the \HGG-
and \HGNS-model values of more than 500,000 asteroids by combining observations
from different
observatories.\footnote{\url{https://wiki.helsinki.fi/display/PSR/Asteroid+absolute+magnitude+and+slope}}
Since the publication of this catalogue, the number of known minor planets has
increased almost two-fold. We aim to extend this effort while taking note of two
caveats of the analysis. First, the fitted \HGG- and \HGNS-model were not
constrained as in \autoref{equ:con_g1andg2}, resulting in 52\% of the reported
slope parameters lying outside the physical range. Furthermore, the authors
combined observations from different wavebands, applying average asteroid
colour-indices to unify the data. However, the slopes and band widths of
asteroid spectra increase with increasing phase angle
\citep[e.g.,][]{10.1006icar.2002.6923,2012Icar..220...36S}, resulting in
wavelength-dependent phase curves. Therefore, we refrain from combining
observations acquired in different wavebands.

Utilising serendipitous observations offers the advantage of large catalogues,
however, the derived phase curves are subject to several undesirable effects.
The majority of observations reported to the \emph{Minor Planet Centre}
(MPC)\,\footnote{\url{https://minorplanetcenter.net}} is collected by
large-scale surveys aiming to monitor the near-Earth environment. To identify
asteroids on collision trajectories with Earth, these surveys favour observing
asteroids in quadrature rather than opposition. This introduces a bias towards
observations at the maximum observable phase angle for asteroid populations with
superior orbits to that of Earth. In addition, the light curve
modulation introduced by rotation and apparition effects can be reduced using
accurate targeted observations, e.g. by means of a Fourier analysis to derive
the shape of the light curve and by treating observations from multiple
oppositions separately. For non-targeted observations, the comparatively large
photometric uncertainty inhibits such a reduction. Furthermore, the
corresponding increase in required observations would decrease the size of the
available sample, diminishing the statistical significance of the resulting
catalogue. As a consequence, serendipitously observed phase curves exhibit 
stochastic fluctuations, translating into larger uncertainties on the fitted
phase coefficients.

In this work, we derive the phase curve parameters of serendipitously observed
asteroids. In \Autoref{sec:methodology}, we describe the observations at hand
and the Bayesian parameter inference approach. The fitted phase curve parameters
are summarized in \Autoref{sec:results}. The taxonomic interpretability of the
\GG-parameters and their wavelength-dependency are outlined in
\Autoref{sec:taxonomy}.  We illustrate these results with the taxonomy of
asteroid families in \Autoref{sec:families}.  In
\Autoref{sec:uncertainty_budget}, we quantify the effect of various sources of
uncertainties and limited phase curve coverage at opposition on the derived
phase curve parameters. The conclusions are presented in
\Autoref{sec:conclusion}. 

%% file: sections/section2.tex
\section{Methodology}%
\label{sec:methodology}
% ------
\subsection{Selection of serendipitous observations}
\label{sub:selection}
The MPC observations database contains 246 million asteroid observations as of
March 2020. We aim to acquire densely sampled phase curves for a large, unbiased
corpus of asteroids. At the same time, we seek to quantify the inherent effects
of the asteroids' shape-induced light curve modulation on the phase curve
parameters. We therefore attempt ~to ~exclude possible sources of systematic
effects rigorously. These derive foremost from non-homogeneous photometry
between different observatories. Differences in the filter transmission,
reduction pipeline, or stellar catalogues introduce discrepancies in the
reported magnitudes of asteroids.

Instead, we choose to utilise observations from a single observatory, maximising
the likelihood of consistent data treatment. In recent years, both the
\emph{Panoramic Survey Telescope and Rapid Response System} (Pan-STARRS,
\citet{10.1002asna.200410300}) and the \emph{Asteroid Terrestrial-impact Last
Alert System} (ATLAS, \citet{10.10881538-3873aabadf}) have placed among the top
five contributors to the MPC in terms of number of observations. Comparing the
ephemerides at the epoch of observation of several thousand asteroids observed
by the surveys, we find that the bias towards observation at asteroid quadrature
is less pronounced in the ATLAS catalogues. In addition, ATLAS has acquired
dual-band photometry of a large number of asteroids at comparable phase angles,
offering an excellent dataset to investigate the wavelength-dependency of the
phase curves. Hence, we make use of observations by ATLAS, referring the reader
to \citet{10.1016j.icarus.2015.08.007} for a derivation of $H,\,G$-parameters
using Pan-STARRS observations.

% ------
\subsection{ATLAS}%
\begin{figure}[t]
  \centering
  \input{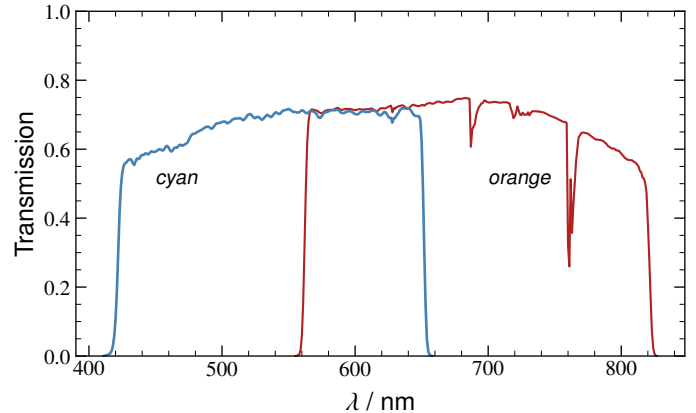}
  \caption{Transmission curves of the \cyan and \orange
    filters used by the ATLAS survey \citep{10.10881538-3873aabadf}. Data from the
    Filter Profile Service of the Spanish Virtual Observatory
  \citep{10.5479adsbib2012ivoa.rept.1015r}.\,\protect\footnotemark}
  \label{fig:atlas_transmission}
\end{figure}

\label{sub:atlas}
% Description ATLAS
ATLAS is a NASA-funded sky-survey aiming to observe \emph{near-Earth asteroids}
(NEAs) on impactor trajectories with the Earth. It was designed with a focus on
a high survey speed per unit cost \citep{10.1086657997}. Two independent 0.5\,m
telescopes located at Haleakala and Mauna Loa in Hawaii are in operation since
2015 and 2017 respectively, achieving multiple scans of the northern sky every
night. Each telescope observes a 30\,deg$^2$ field-of-view. By March 2020, ATLAS
has discovered 426 NEAs, including 44 potentially hazardous
ones.\,\footnote{\url{http://atlas.fallingstar.com}} Standard observations are
carried out in two filters, a bandpass between 420\,-\,650\,nm termed \cyan and
a bandpass between 560\,-\,820\,nm termed \orange. The transmission curves of
these filters are depicted in \Autoref{fig:atlas_transmission}. The observed
asteroid astrometry and photometry are reported to the MPC.

We received dual-band photometry of \numbatlasasteroids distinct asteroids from
the ATLAS collaboration. A third of the objects was observed at phase angles
below 1\,deg. The observations were acquired between June 2015 and December
2018. We extend this database by including ATLAS observations from 2019 reported
to the MPC.

The original database contained 26.8 million observations, to which we add 8
million using the MPC database. The required ephemerides are retrieved using the
IMCCE's \,\,Miriade\,\, tool\,\footnote{\url{http://vo.imcce.fr/webservices/miriade/}}
~~\citep{2008LPICo1405.8374B}. All \numbatlasasteroidso asteroids were observed in
\orange, while \numbatlasasteroidsc were observed in \cyan as well. A small
fraction of visually inspected phase curves showed large outlier magnitudes
likely caused by blended sources in the images. We remove these detections by
rejecting observations where the difference between the predicted and the
observed apparent magnitude was larger than 1\,mag. This cut is well above the
amplitude modulation of asteroid light curves induced by the spin
\citep{10.1016j.pss.2015.06.002,10.10510004-6361201730386}.

% ------
\subsection{Phase parameter inference}%
\label{sub:bayesian_inference_of_phase_parameters}
Fitting scattering model functions to phase curves is notoriously ambiguous and
the results do not necessarily describe the observed surface, especially in the
case of observations where the shape-induced light curve modulation has not been
subtracted \citep{1989A&A...208..320K,10.1016s0019-10350200020-9}. We choose a
computationally expensive Bayesian parameter inference with \emph{Markov ~chain
~Monte Carlo} (MCMC) simulations to fit the ~~\HGG- and \HGS-models, allowing to
examine the posterior distributions of the phase curve parameters. To
differentiate between the absolute magnitudes obtained with these two models, we
use the subscript H$_{12}$ for \HGS.

\footnotetext{\url{http://svo2.cab.inta-csic.es/svo/theory/fps3/}}

For both absolute magnitudes $H$ and $H_{12}$, we choose a weakly informative,
normally distributed prior,
\begin{equation}\label{equ:prior_h}
  p(H), p(H_{12}) = \mathcal{N}(\mu=10, \sigma=100)\,,
\end{equation}
where $\mathcal{N(\mu, \sigma)}$ describes the Gaussian normal distribution with
mean $\mu$ and standard deviation $\sigma$. The mean and standard
deviation are set to approximate a uniform distribution over the relevant
absolute magnitude parameter space. Alternatively, informative prior
distributions could be derived from the distribution of the absolute magnitude
of Main-Belt asteroids, up to the limiting magnitude of ATLAS
\citep[m$\sim$19,][]{10.10881538-3873aabadf}, or from computing least-squares fits
of the $H\,G$-model to each phase curve and using the acquired $H$ and its
uncertainty as moments of the Gaussian distribution.

To quantify the effect of the prior choice, we computed the $H,\,G_1,$ $G_2$- and
\HGS-model fits for 100 randomly chosen phase curves using the three outlined
priors. The distribution of $H$ for Main-Belt objects is approximated with a Gaussian
distribution with $\mu=17.2$ and $\sigma=1.6$. The resulting distributions of
the model parameters $H$ and $H_{12}$ show negligible variation with averaged
differences below 0.01, only the prior based on the $H\,G$-model yields larger
$H$-values with an averaged difference of 0.06 as it limits the size of the
opposition effect. The quantification  supports the choice of the weakly
informative  choice, though the prior based on the Main-Belt magnitudes would have been equally acceptable.

Following the slope parameter constraints in \autoref{equ:constraints}, we
choose uniform distributions between 0 and 1 as prior probabilities for $G_1$,
$G_2$, and $G_{12}^*$,
\begin{equation}\label{equ:prior_g1_g2}
  p(G_1), p(G_{2}), p(G_{12}^*) = \mathcal{U}[0, 1]\,.
\end{equation}
Note that this choice in priors does not necessarily lead to $G_1$ and $G_2$
satisfying constraint~\ref{equ:con_g1andg2}. To accommodate for
this, we remove solutions where $1 - G_1 - G_2 < 0$ from the MCMC samples.

\begin{figure}[t]
  \centering
  \input{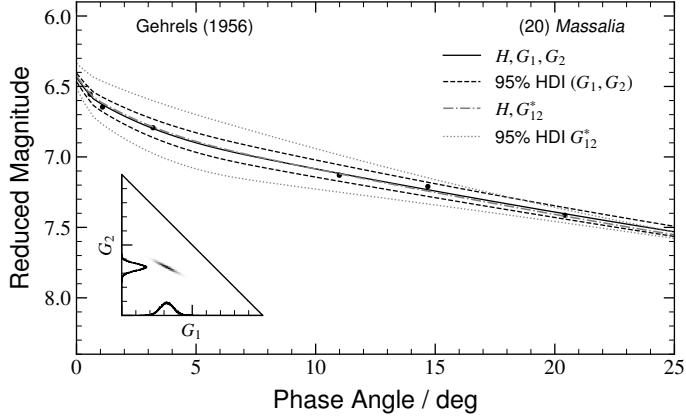}
  \caption{The phase curve of \nuna{20}{Massalia}, as observed by
    \citet{10.1086146166}, fitted with the \HGG-model (solid black). The black dashed curves
    are plotted using the 95\% highest density interval (HDI) values of the three fit
    parameters. The gray, dash-dotted line represents the \HGS-model fit
    with the gray dotted line representing the uncertainty envelope. The
    measurement uncertainties of 0.01\,mag are smaller than the marker size. The
    inset shows the 1\,D- and 2\,D-distributions of the G$_1$ and G$_2$
  Markov chain Monte Carlo samples.}%
  \label{fig:massalia_phase_fit}
\end{figure}

In the following, we collectively describe the parameters of the respective
photometric model using $\boldsymbol{\Theta}$. 

We define the likelihood function by assuming that the observed apparent
magnitudes $m_\alpha$ at phase angle $\alpha_i$ follow a normal distribution
with the true apparent magnitude as mean and a standard distribution
$\sigma_{\alpha,i}$ dictated by the asteroid's light curve modulation and the
observation accuracy,
\begin{equation}
  p(\textbf{m}_\alpha\vert\boldsymbol{\Theta}) =
  \mathcal{N}(\mu=\textbf{m}_{\alpha,i}, \sigma={\sigma_{\alpha,i}})\,.
\end{equation}
With the given prior probabilities, likelihood function and data, the posterior
probability distribution $p(\boldsymbol{\Theta}\vert\textbf{m})$ is defined.
However, we cannot derive it analytically and need to approximate it by means of
MCMC simulations.

We use the \texttt{pymc3} \texttt{python} package
\citep{10.7717peerj-cs.55}\,\footnote{\url{https://docs.pymc.io/}} to perform
these simulations. The photometric models are implemented in the \texttt{sbpy}
package
\citep{10.21105joss.01426}\,\footnote{\url{https://sbpy.readthedocs.io/}}. As
best-fit parameters, we use the mean values of the respective parameter's
posterior probability distribution. The uncertainties are given by the bounds of
the 95\% \emph{highest density interval} (HDI) of the posterior distributions.

\begin{figure}[t]
  \centering
  \input{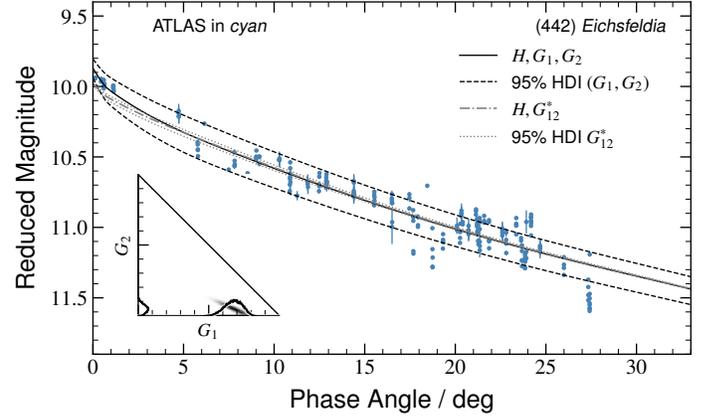}
  \caption{Like \Cref{fig:massalia_phase_fit}, using the non-targeted ATLAS
    observations of \nuna{442}{Eichsfeldia} in \cyan instead. The \HGS-model
    deviates towards the opposition effect as the \GG-parameters of the asteroid are
    outside the definition of \GS. This further leads to the unreasonably
  small 95\% highest density interval of the \GS parameter.}
  \label{fig:eichsfeldia_phase_fit}
\end{figure}

In \Autoref{fig:massalia_phase_fit}, we depict the parameter inference for the
phase curve of \nuna{20}{Massalia}, as observed by \citet{10.1086146166}, who
first noted the opposition effect on the surface of an asteroid using these
targeted observations. The resulting \HGG- and \HGS-model fits are displayed,
including the 1\,D- and 2\,D distributions of the $G_1$ and $G_2$ MCMC samples.
The joint distribution of \GG illustrates that the uncertainty in the fit
derives primarily from the photometric slope $k$ as opposed to the size of the
opposition effect, as can be seen in the uncertainty profile of the fitted phase
curve (dashed lines). Nevertheless, the posterior
distributions of $G_1$ and $G_2$ are Gaussians with comparatively small standard
deviations due to the targeted nature of the observations.

In comparison, we show the effect of serendipitous observations on the parameter
inference in \Autoref{fig:eichsfeldia_phase_fit}, which depicts the $H,\,G_1,$ $G_2$- and
\HGS-model fits to observations of \nuna{442}{Eichsfeldia} by ATLAS in \cyan.
The same range of reduced magnitudes is given on the y-axis as in
\Autoref{fig:massalia_phase_fit}. The light curve modulation yields a large
dispersion in reduced magnitude at each phase angle, which is reflected in the
dispersion of MCMC \GG-samples, depicted in the plot inset. We further observe
that the posterior distribution tends towards the unphysical \mbox{$G_2\,<\,0$} -regime,
likely due to the step in reduced magnitude around 27\,deg phase angle.

\Autoref{fig:eichsfeldia_phase_fit} shows that the two photometric models arrive
at similar slopes, while the size of the opposition effect and the inferred
absolute magnitudes vary considerably ($H=9.87$, $H_{12}=9.98$). This highlights
the restricted nature of the \HGS-model; the \GG-parameters of
\nuna{442}{Eichsfeldia} in \cyan are (0.64, 0.05), which is distant from the
linear \GG-relation of \GS. Hence, the \HGS-model cannot adequately describe the
opposition effect of the phase curve. 

%% file: sections/section3.tex
\section{Results}%
\label{sec:results}

In this section, we present the phase curve parameters acquired by applying the
\HGG- and \HGS-models to the serendipitous ATLAS observations.
\citet{Erasmus_2020} investigate the taxonomic interpretability of the
\cyan-\orange colour to identify asteroid family members. We highlight the
differences in the absolute magnitudes derived with the \HGG- and \HGS-model.

\subsection{Sample selection and data availability}%
\label{sub:sample_selection}

From the \numbatlasasteroids asteroids observed by ATLAS, we select the ones
with at least one observation at a phase angle $\alpha$ below three degree,
$\alpha_{\mathrm{min}}\leq\phminlimit$\,deg, to ensure an adequate description
of the opposition effect. This decreases the sample size to \numbtwodeg
asteroids, rejecting almost a third of the available sample. We choose this
limit based on the significant importance of the opposition effect on the phase
curve parameters, and after simulating different degrees of incomplete phase
curve coverage towards opposition, refer to \Autoref{sec:uncertainty_budget}.
We further apply lower limits on the number of observations, $N\geq\Nlimit$, and
the maximum phase angle of observation, $\alpha_{\mathrm{max}}\geq10$\,deg, to
remove sparsely-sampled phase curves. The final sample consists of
\numbuniqueasteroidswithphasecurve unique asteroids, \numbphasecurvesc observed
in \cyan and \numbphasecurveso observed in \orange.

We provide the \HGG- and \HGS-model parameters for all \numbphasecurves fitted
phase curves in an online catalogue\ifx\destination\arxiv
\,\footnote{The catalogue is
    available at the CDS via anonymous ftp to \url{cdsarc.u-strasbg.fr}
    (\url{130.79.128.5}) or via \url{http://cdsarc.u-strasbg.fr/viz-bin/qcat?VII/288}
  }
\fi publicly available at the \emph{Centre de Données
astronomiques de Strasbourg}
(CDS).\,\footnote{\url{http://cdsweb.u-strasbg.fr/}} The format of the catalogue
is described in \ref{app:online_catalgoue}. For \numbfailedfits phase curves,
the \HGG-model failed to fit the phase curve, meaning that not a single
of the 48,000 MCMC samples satisfied \autoref{equ:constraints}. By visually
inspection, we found that large magnitude dispersions, insufficient sampling, or
strong apparitions effects lead to unphysical shapes of the phase curves, where
the magnitude decreased with increasing phase angle.  An example is given in
\Autoref{fig:bettina_apparitions} in \Autoref{sec:uncertainty_budget} for the
phase curve of \nuna{250}{Bettina}, exhibiting a particularly strong apparition
effect.

\begin{figure}[t]
  \centering
  \input{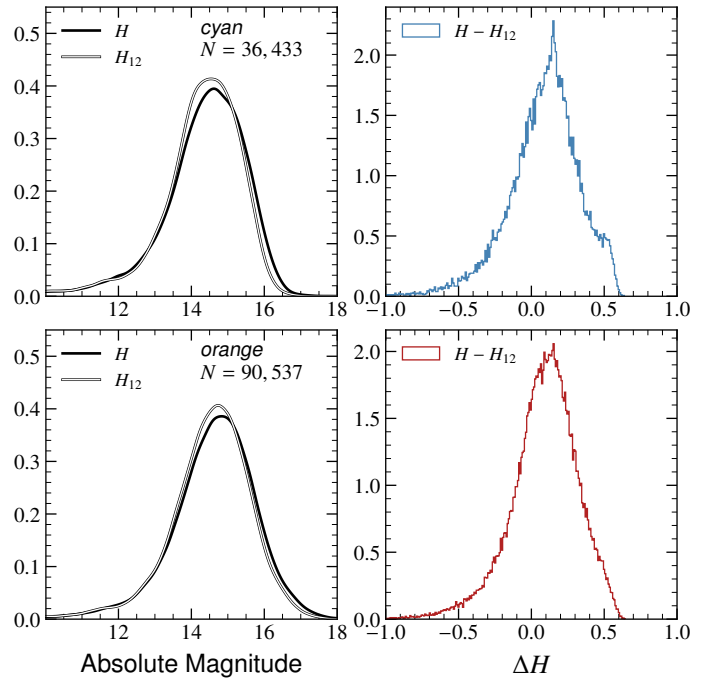}
  \caption{\textit{Left:} The distribution of absolute magnitudes $H$ (black)
    and $H_{12}$ (white) derived from ATLAS phase curves of
    \numbuniqueasteroidswithphasecurve asteroids using the \HGG- and \HGS-models
    respectively, for phase curves observed in \cyan (top) and \orange (bottom). For
    readability, magnitudes below 10 (\percbelowtenmag\% of the sample) are not
    shown. \textit{Right:} The difference in the absolute magnitude derived with the
  two models, for phase curves in \cyan (top) and \orange (bottom).}
  \label{fig:h_h12_delta}
\end{figure}

\subsection{Phase curve parameters in cyan and orange}%
\label{sub:absolute_magnitudes_in_cyan_and_orange}

We display the absolute magnitudes $H$ and $H_{12}$ in \cyan and \orange derived
from the model fits on the left hand side of \Autoref{fig:h_h12_delta}. It is
apparent that the absolute magnitudes from the \HGS-model are lower on average.
This is highlighted in the right hand side part of \Autoref{fig:h_h12_delta},
where we display the histograms of the difference $H - H_{12}$ for all objects.
Both in \cyan and in \orange, the distributions peak around 0.1\,mag and extend
up to 1\,mag absolute difference.

\begin{figure}[t]
  \centering
  \input{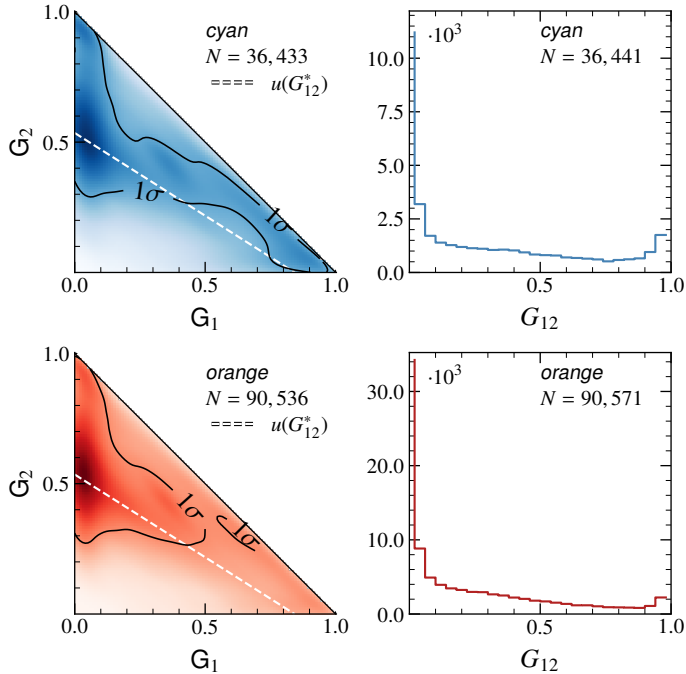}
  \caption{\textit{Left:} The 2\,D-KDE distribution fitted to the \GG
    parameters of the phase curves observed in \cyan (top) and \orange (bottom).
    The black contours outline the 1\,$\sigma$-levels of the KDE distributions.
    \textit{Right:} The histogram of the \GS-parameter derived from the same sample
  of phase curves, aligned in the same order.}
  \label{fig:g1g2all}
\end{figure}

The origin of these discrepancies can be seen in \Autoref{fig:g1g2all}. On the
left hand side, we give the 2\,D \emph{kernel density estimator} (KDE)
distribution fitted to the \GG-pairs of the whole sample in \cyan and \orange
using a Gaussian kernel. The black 1\,$\sigma$-contour gives the KDE level at
which 68\% of the summed probabilities is contained in the area, resembling the
1\,$\sigma$-level of a Gaussian distribution. The \GS-parameter space is
superimposed as white, dashed line (refer to \Autoref{equ:g12star}). We observe
a clustering towards low $G_1$, medium $G_2$ values in both wavebands, centered
around the region where we expect S-type asteroids to be located (refer to
\Autoref{sec:taxonomy}). S-types dominate the inner and middle Main Belt in
terms of absolute number \citep{10.1016j.icarus.2013.06.027}. The distributions
further extend towards
larger photometric slopes to the region of low-albedo complexes such as the
C-types, with a larger fraction of low-albedo asteroids visible in the
\cyan-band. Further noticeable is an extension of distributions towards $G_2=1$,
i.e.\,negligible photometric slopes and opposition effects, indicative of
high-albedo complexes.

In both wavebands, the majority of asteroids exhibits \GG-values above the
\GS-definition in \GG-space. The \HGS-model therefore fails at describing these
phase curves, particularly the size of the opposition effect will be
overestimated for objects above the \GS-line. $\zeta-1$ increases non-linearly
towards the origin of \GG-space, leading to the large tail towards negative
differences of the distributions on the right hand side of
\Autoref{fig:h_h12_delta}.

The \GS-parameters of the phase curves are depicted in the histograms on the
right hand side of \Autoref{fig:g1g2all}. \gscbelow\% of the sample in \cyan and
\gsobelow\% of the sample in \orange are below 0.1, resembling closely the
distribution in \GG-space. In both bands, we observe a decline of the number of
objects towards larger \GS values up to about 0.9, where the number rises again.
These tendencies of \GS towards the limiting 0 and 1 values indicate that a
large number of phase curve lies outside the defining relation, hence, they
cannot be represented appropriately by the model. We point out that we observe
phase curves fitted with \HGG on the edges of \GG-space as well, though in a
much smaller ratio. As mentioned in \Autoref{sec:methodology}, we attribute
these to stochastic magnitude variations leading to unphysical shapes of the
observed phase curves.

\subsection{Suitability of \HGS for taxonomic classification using non-targeted
observations}

The different results between \HGG and \HGS are expected as the \HGG is more
flexible due to the third photometric parameter.
\citet{10.1016j.icarus.2010.04.003} stress that the main advantage of the
\HGNS-model with its reduced parameter space is its predictive power when
utilized with sparsely-sampled phase curves. Indeed, giving
non-targeted, sparse observations, and a prior knowledge on the target
taxonomy, \citet{10.1016j.pss.2015.08.010} show that the absolute magnitude can
be estimated using class-specific fixed slope parameters in the fitting
procedure.

However, regarding a taxonomic classification based on the parameters of the
\HGS-model, we conclude here that neither the absolute magnitudes nor the slope
parameter are sufficiently reliable. The discrepancy between $H$ and $H_{12}$
prevents classification based on the absolute magnitude. To compare, we compute
the colours of the asteroid taxonomic classes in the \emph{Sloan Digital Sky
Survey} (SDSS)\,\footnote{\url{https://www.sdss.org/}} using the spectral
templates of the classes from
\citet{10.1016j.icarus.2013.06.027}.\,\footnote{The template spectra are
retrieved from \url{http://smass.mit.edu/busdemeoclass.html}.} For each colour,
we compute the average difference between the complexes, resulting in
\meandiffug\,mag ($u$-$g$), \meandiffgr\,mag ($g$-$r$), \meandiffri\,mag
($r$-$i$), and \meandiffiz\,mag ($i$-$z$). The inaccuracies introduced by the
\HGS~-~model are on average greater than these differences, preventing taxonomic
classification. As outlined in
\Autoref{sub:absolute_magnitudes_in_cyan_and_orange}, we regard the
\GS-parameter insufficient for any conclusion on the surface composition as
well.

%% file: sections/section4.tex
\section{Taxonomy} \label{sec:taxonomy}

In the following, we evaluate the taxonomic information content of the phase
curve parameters, focusing on \GG-values derived from the serendipitous phase
curves. We illustrate the distributions of the asteroid complexes and quantify
their similarities in \cyan and \orange. Further evaluated are their
wavelength-dependency and the ability to solve degeneracies of asteroid spectra
using phase curve parameters. 

The \GG- and $G_{12}$-distributions of different complexes have been studied by
\citet{10.1016j.jqsrt.2011.03.003} and\citet{10.1016j.pss.2015.11.007}. We do
not further explore the \GS-parameter following the conclusion of
\Autoref{sec:results}.

\subsection{Importance of observing the opposition effect}
\label{sub:ioe}

In a first iteration, we performed the following analysis on all acquired phase
curves, using the limits on $\alpha_{\mathrm{min}}$, $\alpha_{\mathrm{max}}$,
and $N$ as outlined in \Autoref{sub:sample_selection}. However, we noticed large
dispersions in the arising \GG-distributions of the complexes, which showed a
clear trend with respect to the number $N$ of observations in each phase curve.
\GG-parameters derived from phase curves with low $N$ dispersed more from the
center of the distributions than the ones from more densely covered phase
curves. 

The vital role of the opposition effect both for determining the absolute
magnitude $H$ and the taxonomic interpretation of the phase curve has been
pointed out in the previous sections. Its non-linear dependence on the phase
angle and the inherent magnitude dispersion of the serendipitous observations
(refer to \Autoref{sec:uncertainty_budget}) require a dense coverage of
observations to accurately describe the brightness surge. As ATLAS aims to
observe asteroids on impact trajectory, only 7.3\% of the 24 million
observations analysed here have been acquired of asteroids at $\alpha\leq3$deg,
i.e. close to opposition, see \Autoref{fig:phase_angle_dist}. For $N=50$, this
corresponds to 3-4 observations covering the most important part of the phase
curve.
We therefore evaluated the trade off between dispersion introduced in \GG-space
by phase curves with insufficient sampling of the opposition effect and by small
sample numbers in less common asteroid taxa. Through visual inspection of the
resulting complex distributions, we settled on $N=125$ as limit for the
following analyses, decreasing the initial sample size of \numbphasecurves by
more than half, down to \numbphasecurvestaxlimit. We stress that this large
number of required observations stems from the science goal of the observatory
providing the data; future large scale surveys like LSST can derive accurate
phase curves from fewer observations provided the opposition effect is in the
focus of the observation schedule.

\subsection{Complex mapping} \label{sub:complex_selection}

We retrieve previous taxonomic classifications from various references for
\asteroidswithclasstaxlimit objects, in addition to reference albedo values for
\asteroidswithalbedotaxlimit of these classified asteroids. The albedos are
employed to identify misclassifications and to separate classes into different
complexes as outlined below. We collected these values from numerous sources and
refer the reader to \ref{app:online_catalgoue} and the online catalogue of
the phase curve parameters for details.

The majority of classifications follow the Bus- or Bus-DeMeo-schemes
\citep{10.1006icar.2002.6856,10.1016j.icarus.2009.02.005}, which are performed
on low-resolution asteroid reflectance spectra. As the phase curve parameters
\textsl{a priori} contain less taxonomic information than spectra and to
increase the size of the subsamples, we map the classes into broader taxonomic
complexes. In the Bus-DeMeo taxonomy, there are 25 classes spanning a space of
13 complexes which are designated by unique letters of the alphabet
\citep{10.1016j.icarus.2018.12.035}. We map the asteroids onto these complexes
based on their previous classifications. For classes which have been defined in
previous taxonomies but are no longer present in the Bus-DeMeo one, we choose
the current complex resembling the previous class the most. As an example, the
F-type defined in \citet{1984PhDT.........3T} is mapped onto the B-complex.

Previous taxonomies like \citet{10.1086115007} differentiate the X-type
asteroids into low-albedo P-types, medium-albedo M-types, and high-albedo
E-types. Asteroids with the same spectral shape but lacking albedo measurement
are grouped into the X-types. As the albedo was dropped in subsequent
taxonomies, so was the differentiation of the X-type classes. Given the
correlation of the phase curve parameters with the albedo
\citep[e.g.,][]{10.1006icar.2000.6410,10.1016j.pss.2015.08.010,2018EPSC...12..730B}, we expect the
X-type asteroids to separate in \GG-space. Therefore, we map asteroids
classified in the X-complex into the P- ($p_V \leq 0.075$), M- ($0.075 < p_V <
0.30$), E- ($p_V \geq 0.30$), and X-complex (no reference albedo available),
following the limits in \citet{1984PhDT.........3T}.

Hydrated C-types make up more than 30\% of C-types in the Main Belt
\citep{10.1016j.icarus.2012.08.042,10.1016j.icarus.2014.01.040}. The aqueous
alteration is imprinted in absorption bands at 0.7- and 3.0\,$\mu$m. The ATLAS
\orange filter covers the 0.7\,$\mu$m band, therefore, these classes may
separate in phase-parameter space. We split the Cgh- and Ch-types from the
C-complex to investigate whether \GG reveal the hydration.

The final mapping of classes to complexes is given in
\Autoref{tab:complex_mapping}. Due to the low number of O-type asteroids, we
exclude the complex from the analysis. We further rejected several ambiguous
class assignments such as DS, CQ, SA, CS, XS from
\citet{10.10510004-6361200913322}, which were performed on low-resolution
visible photometry from the SDSS and given to objects which presented
photospectra with different features in different observations.  Further, the
D-complex contains more than 200 objects with albedos between 0.1 and 0.5,
indicating that they are misclassified. We therefore introduce an upper limit of
0.1 albedo on the D-type complex. 

Finally, we exclude the Ad, Bk, Ds, and Kl classes from
\citet{10.10510004-6361201833023} temporarily. These classes are assigned based
on near-infrared spectrophotometry using the VISTA-MOVIS catalogue
\citep{2016A&A...591A.115P}. The spectra of these types are degenerate in the
regarded wavelength regime, therefore, the objects are classed together. The
authors note that these classes are likely made up objects belonging to the
denominating complexes (i.e.\,Ad is made up of A- and D-type asteroids). In a
subsequent analysis step, we investigate the class ratios in these combinations
using the phase curve parameters. We choose the VISTA-MOVIS sample rather than
the SDSS sample by \citet{10.10510004-6361200913322} as the degeneracy in
near-infrared cannot be resolved without additional information such as the
phase curve parameters. For the ambiguous SDSS results, additional observations
in the visible could suffice to resolve the classifications.

\begin{table}[t]
  \centering
  \caption{The applied mapping of asteroid taxa to complexes. The previous
    classifications are mapped to the complex denoted under $\Sigma$. $N$ refers to
    the number of asteroids in each complex. $\bar{p_V}$ and $\sigma_{p_V}$ give the
    mean visual albedo and its standard deviation respectively of all asteroids in
    the complex. The X-complex does contain asteroids with albedo measurements by
  definition.}
  \input{tables/complex_mapping.tex}
  \label{tab:complex_mapping}
\end{table}

\begin{figure*}[pt]
  \centering
  \input{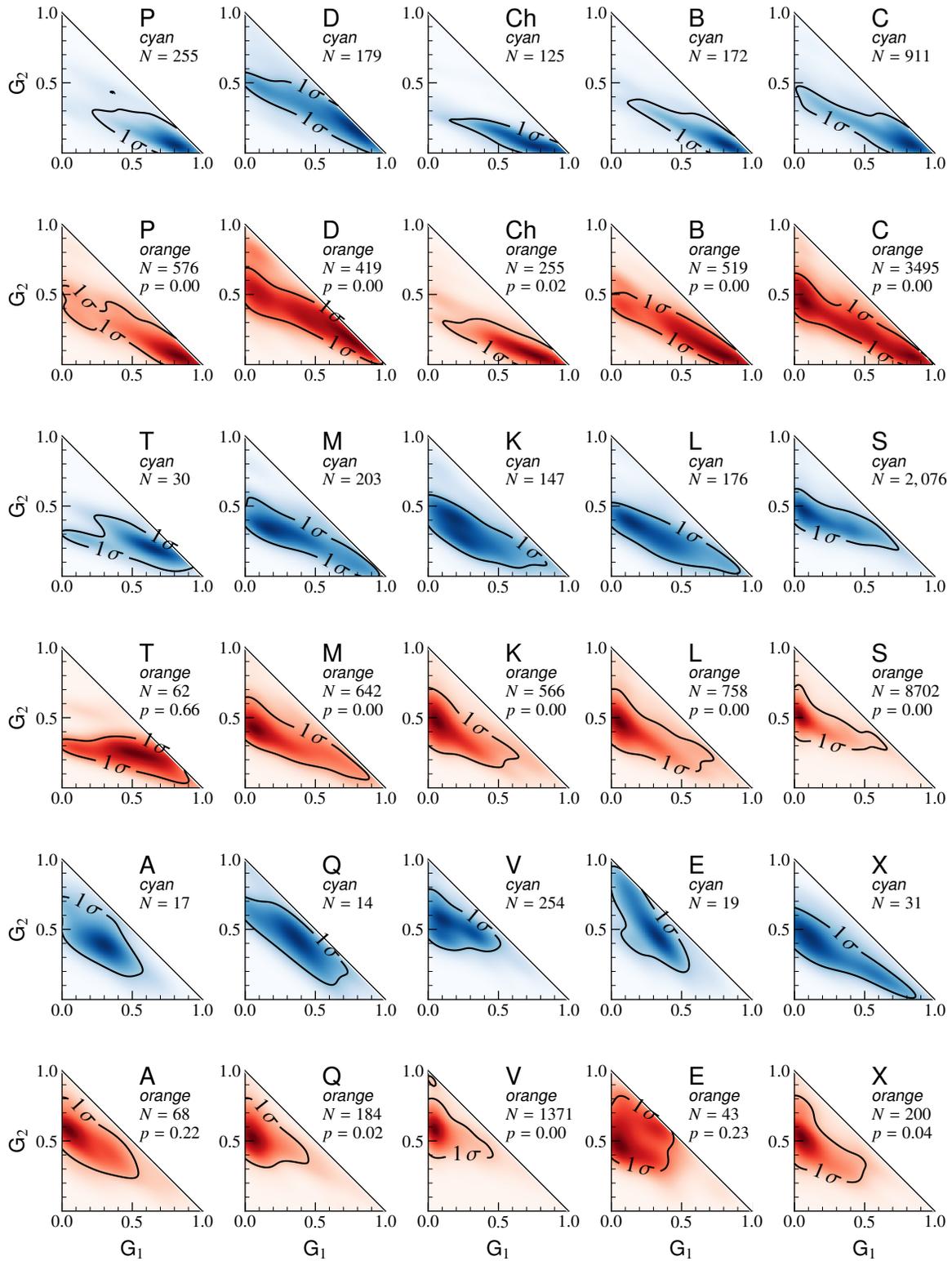}
  \caption{Depicted are the \GG-distributions for several taxonomic asteroid
    complexes comprising \asteroidswithclasstaxlimit objects, derived from
    serendipitous phase curves observed by ATLAS in \cyan and \orange.  The
    complexes are sorted in increasing order of their average visual albedo. The
    distributions are represented by 2\,D Gaussian \emph{kernel density estimators}
    (KDE) fitted to the \GG-pairs. The black contours give the KDE level at which
    68\% of the summed probabilities is encompassed, resembling the
    1\,$\sigma$-level of a Gaussian distribution. Further given are the number of
    asteroids $N$ in each complex and waveband as well as the two-sample 2\,D
    Kolmogorov-Smirnov $p$-values computed between the distributions in \cyan and
  \orange for each complex.} 
  \label{fig:g1g2_taxonomy_all}
\end{figure*}    

\begin{table*}[t]
  \centering
  \caption{For each complex, we provide the number $N$ of analysed phase curves
    as well as the geometric center $C$ and area $A$ of the 95\%-probability contour
    in \cyan (subscript $c$) and \orange (subscript $o$).  The areas are multiplied
    by 1,000 for notation purposes. Further given are the photometric slope
    parameter $k$, the size of the opposition effect $\zeta-1$, and the
    Kolmogorov-Smirnov $p$-values for each complex in this study. $k$ and $\zeta-1$
    are calculated using the \GG-pairs of the geometric centers, following
  \autoref{equ:k} and \autoref{equ:z}.}
  \input{tables/complexes.tex}
  \label{tab:complexes}
\end{table*}

\subsection{Distribution of taxonomic complexes in \GG--space}%
\label{sub:distribution_of_taxonomic_complexes_in_gg_space}

The 2\,D KDE distributions fitted to the \GG-parameters of the phase curves of
the 15 complexes are shown in \Autoref{fig:g1g2_taxonomy_all}, both in \cyan and
\orange, with a black contour marking the KDE level at which 68\% of the summed
probability is encompassed. The complexes are depicted in increasing order of
their average visual albedo. It is readily apparent that the albedo-dependence
of the opposition effect and photometric slope as described by
\citet{10.1006icar.2000.6410} is present in the ATLAS observations as well; with
increasing average visual albedo, the distribution centers shift from large
$G_1$-values towards medium- and finally large $G_2$ values, i.e.\,towards
flatter phase curves and smaller opposition effects. We further find good
agreement with the \GG~-~parameters extracted from targeted campaigns by
\citet{10.1016j.pss.2015.11.007} and \citet{10.1016j.pss.2015.08.010}. The
medium- and high-albedo S-, M-, and E-types populate regions of small
photometric slopes, while low-albedo B-, C-, D-, and P-types present much larger
slopes. Overall, the intermittent region around $G_1$=0.5 is sparsely populated;
only the K- and T-complexes in both wavebands and the \mbox{L-,} and M-complexes
in \cyan present large probabilities there. We summarize the distributions in
\Autoref{tab:complexes}, giving the \GG-coordinates of the geometric center of
the 95\%-probability-level contour for each complex.  Further stated are the
sizes of the areas encompassed by the 95\%~-~probability contours, approximating
the dispersion of the complexes in \GG-space after outlier rejection.

The strong disparity in the distributions of the E- and P-complexes shows that
the phase coefficients present a reliable distinction between members of the
X-complex, independent on reference albedo measurements. For the complexes where
we discriminate based on on albedo, i.e.\, the P-, M-, \mbox{E-,} and D-type, we
see large tails in the 1\,$\sigma$-distributions, which we attribute to
remaining misclassifications. Asteroid albedo measurements carry uncertainties
around 17.5\% \citep{10.38471538-3881aacbd4}, suggesting that the P-, M-, and
E-complexes are overlapping due to these interlopers. The C- and D-types present
broad distributions, specifically in the \orange samples. This indicates a
substantial fraction of misclassifications in the literature. The majority of
classifications is retrieved from visible photometry based on the SDSS. As noted
in \Autoref{sub:complex_selection} and \citet{10.10510004-6361200913322},
asteroids can display ambiguous spectral features of several taxonomies, leading
to mixing of high- and low-albedo classifications (misclassification of X- to
C-types and S- to D-types). This hypothesis is further supported by the
distribution of the Ch-complex. The classification of hydrated C-types is
subject to more scrutiny than the more general C-types, hence we expect a much
smaller fraction of misclassifications.  Indeed, we observe less dispersed
\GG-distributions in the lower-albedo regime for the Ch-complex. Finally, the
contamination of the C-complex prevents a conclusion on the ability to observe
hydration in slope parameter space.

We conclude that the parameters of phase curves carry substantial taxonomic
information, even for serendipitously acquired observations. Several
observational requirements need to be fulfilled, such as a dense coverage of the
opposition effect. Nevertheless, this promises a classification dimension as
insightful as the albedo while being more accessible to the observer.

\subsection{Wavelength-dependency}%
\label{subsec:wavelength_dependency}

Phase reddening describes the steepening of the spectral slope and a change in
the bandwidths of asteroid spectra with increasing phase angle. The effect is
non-linear, see \citet{2012Icar..220...36S}. As the asteroid spectra are
phase-angle dependent, it follows that their phase curves in turn are
wavelength-dependent, resulting in varying \GG~-parameters.
\citet{2015A&A...580A..98C} investigate the phase-angle dependency of taxonomic
classifications of asteroids in the visible wavelength-regime. They find that
the taxonomic complexes are affected to different degrees; objects presenting
the 1\,$\mu$m-olivine/pyroxene-band show stronger correlations between spectral
slope and phase angle than asteroids lacking the absorption feature.

The wavelength-dependency of the \GG-parameters is underlying to the question of
whether it is admissible to combine observations acquired in different
wavelength-regimes to overcome incomplete phase curve coverage. Though the
overlap of the ATLAS \cyan and \orange filters decreases the apparent
wavelength-dependency (refer to \Autoref{fig:atlas_transmission}), the dataset
at hand offers a prime opportunity to investigate the dependency using similar
asteroid samples, phase curve coverages, and apparent magnitude reduction
pipelines.

\begin{figure}[t]
  \centering
  \input{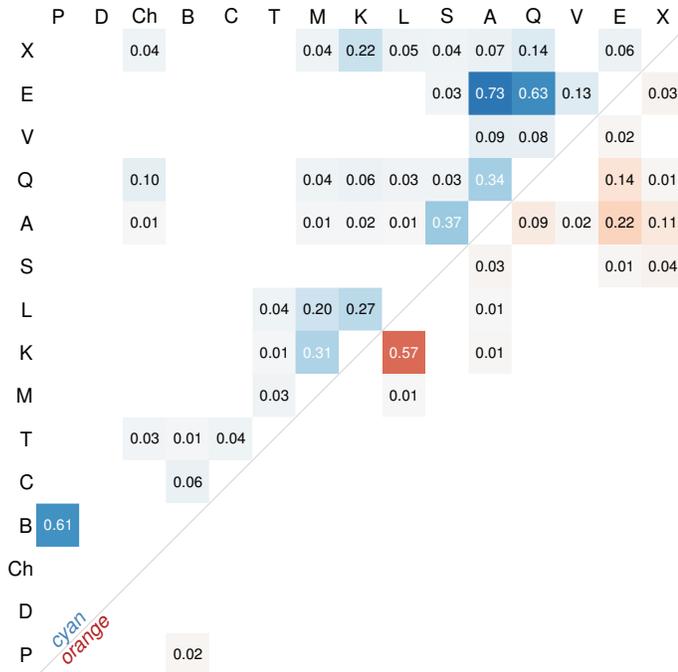}
  \caption{The two-sample 2\,D Kolmogorov-Smirnov $p$-values estimating the
    similarity between the observed \GG-pair distributions of the 15 asteroid
    taxonomic complexes in \cyan (upper left) and \orange (lower right). Values
    above 0.2 indicate that the two paired complexes may have the same underlying
  distribution in \GG-space. Values below 0.01 are not shown for readability.}
  \label{fig:g1g2_taxonomy_heatmap}
\end{figure}    

\begin{figure*}[t]
  \centering
  \input{gfx/g1g2_taxonomy_vista.pgf}
  \caption{Depicted are the \GG-distributions for four spectral classes from
    \citet{10.10510004-6361201833023} containing two distinct asteroid taxa. The
    2\,D kernel density estimates of their \GG-distributions are shown for
    observations in \cyan (blue) and \orange (red). The black contour gives the
    1\,$\sigma$-level. The white letters denote the position of the asteroid
    complexes in \GG-space. $N$ gives the number of asteroids in each sample, while
  the derived ratios of the principal classes are given below.}
  \label{fig:g1g2_taxonomy_vista}
\end{figure*}

We regard the \GG-distributions acquired in \cyan and \orange as two independent
samples. The \emph{Kolmogorov-Smirnov} (KS) $p$-value statistic evaluates the
probability of the null hypothesis that the underlying distribution of the two
compared samples is identical \citep{10.1093mnras202.3.615}. In general,
provided the two compared samples are sufficiently large,
$p$-values above 0.2 indicate a strong similarity, while values below 0.2 reject
the null hypothesis. The results are given in \Autoref{fig:g1g2_taxonomy_all}
above the \GG-distribution \orange of each complex as well as in
\Autoref{tab:complexes}. Most complexes present $p$-values equal or close to
zero, i.e.\,they show wavelength-dependency. The A-, E-, and T-complex are above
the 0.2-threshold. They are the three smallest samples, however, and their
\GG-distributions are noticeably different. We therefore conclude that combining
observations acquired in different wavebands should be strictly avoided.
Additional support for the wavelength dependency of the phase curves
can be derived from the Euclidean distance of the \GG-pairs in \cyan and \orange
for objects observed in both bands. Computing the distances yields a
distribution with mode at 0.127, in good agreement with the displacements of
the complex centroid centers between the two wavebands given in
\Autoref{tab:complexes}.

Differences in the slopes of the phase curves observed at different effective
wavelengths lead to spectral reddening which is proportional to the phase angle
of observation. This is of particular importance for the ESA \emph{Gaia}
mission, which is scheduled to release asteroid spectra obtained at large solar
elongation in its third data release in 2021 \citep{10.1016j.pss.2012.07.029}.
The acquired \GG-distributions describing the shapes of the phase curves allows
us to quantify the amount of spectral reddening per degree phase angle for each
taxonomic complex between the effective wavelengths of the \cyan and \orange
bands. The spectral slope in units of \%/100\,nm is given by
\begin{equation}%
  \label{equ:spectral_slope}
  S_S = \frac{f_o-f_c}{\lambda_o - \lambda_c} \cdot 10^4\,,
\end{equation}
where $f_c$ and $f_o$ are the observed reflectance in \cyan and \orange, and
$\lambda_c=518\,$nm and $\lambda_o=663\,$nm are the effective wavelengths. By
relating the reflectances to the apparent magnitudes using the Pogson scale, we
can express the spectral slope as
\begin{equation}%
  \label{equ:spectral_slope_rewritten}
  S_S = \frac{f_c (10^{-0.4(m_o-m_c)}-1)}{\lambda_o - \lambda_c} \cdot 10^4\,.
\end{equation}
Normalizing the reflectance at $\lambda_c$ gives $f_c=1$, and the remaining
variable is the difference $m_o-m_c$, which we can derive using the
phase curves $m_c(\alpha)$ and $m_o(\alpha)$,
\begin{equation}%
  \label{equ:deltam}
  \Delta m = m_o(\alpha, H_o, G_{1,o}, G_{2,o}) - m_c(\alpha, H_c, G_{1,c},
  G_{2,c})\,.
\end{equation}

\subsection{Identification of interlopers with \GG}%
\label{subsec:identification_interlopers_gg}

The \GG-parameters offer an additional dimension to taxonomic classification,
which is predominantly done in spectral space. The combination of both
dimensions allows to identify interlopers and misclassifications.

Using the 2\,D- KS statistic, we compute the $p$-values to quantify the
resemblance of the asteroid taxa in \GG-space. In
\Autoref{fig:g1g2_taxonomy_heatmap}, we display the heatmap of the two-sample
2\,D KS $p$-values quantifying the similarity of the distributions. The
intersections on the upper left hand side compare the distributions in \cyan,
while the \orange waveband comparison is depicted on the lower right hand side.
The average visual albedo increases towards the upper right.
Complex-combinations yielding $p$-values below 0.01 are left blank for
readability.

Two trends are visible in the heatmap. First, high-albedo complexes tend to show
more resemblance to each other than low-albedo complexes, where only the P- and
B-complexes in \cyan show strong likeness. Second, the complexes present larger
$p$-values in \cyan, where nine pairings cross the 0.2-threshold, prohibiting
the rejection of the null hypothesis, as opposed to two pairs in \orange. Both
pairs, the K-, L- and the A-, E-complexes, cannot be distinguished in either
waveband.

The degeneracies in phase curve parameter space appears reversed to the
degeneracies in spectral feature space. High-albedo objects depict distinct
absorption band properties in band depth, width, and wavelength, which allows
for differentiation even in low-resolution data. Low-albedo types are separated
based on their spectral slopes, which is in general less certain
\citep{2020arXiv200405158M}. The phase parameters offer a complimentary
classification space.

We apply this conclusion to four classes reported by
\citet{10.10510004-6361201833023} in the VISTA-MOVIS based classification. As
outlined in \autoref{sub:complex_selection}, the near-infrared photometry
presents several degenerate classes, of which we show the \GG-pairs in
\Autoref{fig:g1g2_taxonomy_vista}. To estimate the ratios of the different taxa,
we compute the distance in \GG-space for each object to the center coordinates
of the complexes and assign the object to the complex it is closer to. This is a
simple test and proper interloper identification should be performed accounting
for the complete complex distributions; nevertheless, it is used as a proof of
concept here. As we are working with center coordinates derived from statistical
ensembles, we may misclassify single objects. However, the derived probabilities
should hold for the entire samples.

The resolution of degeneracies is effective for the classes at opposite ends of
the albedo spectrum, which here are the A-D- and D-S combinations. We retrieve
the same ratios for both wavebands in these combinations, three-quarters of
A-types in the former and about two-thirds of S-types in the latter.

For the Bk superposition, we observe almost identical ratios as well, while we
note that the observed \GG-distribution peaks between the two complex centers.
Properly accounting for the dispersion of the complexes in \GG-space might
change the retrieved ratios considerably. The Kl class cannot be resolved as
expected following \Autoref{fig:g1g2_taxonomy_heatmap}.

Thus, we conclude that \GG-values derived from serendipitous observations are
sufficient to untangle degeneracies arising in spectral feature space if the
classes separate in albedo-space. Lower-albedo classes may even be separated
from one another provided a reliable observation of the size of the opposition
effect, which is the principal distinction between the B-, C-, D-, and P-types
in \GG-space.

\subsection{Distribution of taxonomic complexes in \GS--space}%
\label{sub:distribution_of_taxonomic_complexes_in_gs_space}

\citet{10.1016j.icarus.2012.02.028} found the S-, C-, and X-types follow
Gaussian distributions in $G_{12}$-space (rather than \GS-space). Following the
discussion in \Autoref{sec:results}, we do not expect any reliable taxonomic
information in the \GS-distributions. However, for completeness, we show them
analogously to \Autoref{fig:g1g2_taxonomy_all} in \Autoref{fig:g12all}. Note
that by definition, the D-, E-, and P-complexes cannot be modelled with \HGS,
hence, we use a dashed linestyle for their distributions.

%% file: tables/complex_mapping.tex
\begin{tabular}{lclrrr}	\toprule
	Class & & $\Sigma$ & N & $\bar{p_V}$ & $\sigma_{p_V}$\\
	\midrule
	P, PC, PD, X, XC,\\XD, XL, Xc, Xe, Xk, Xt& $\rightarrow$ & P & 593 & 0.05 & 0.02 \\
	D, DP & $\rightarrow$ & D & 425 & 0.06 & 0.02 \\
	Cgh, Ch & $\rightarrow$ & Ch & 266 & 0.06 & 0.06 \\
	B, F, FC & $\rightarrow$ & B & 523 & 0.08 & 0.06 \\
	C, CB, CD, CF, CG,\\CL, CO, Cb, Cg, Cgx,\\Co & $\rightarrow$ & C & 3,670 & 0.09 & 0.09 \\
	T & $\rightarrow$ & T & 62 & 0.12 & 0.06 \\
	M, X, XD, XL, Xc,\\Xe, Xk, Xt& $\rightarrow$ & M & 660 & 0.15 & 0.05 \\
	K & $\rightarrow$ & K & 586 & 0.18 & 0.09 \\
	L, LQ, Ld & $\rightarrow$ & L & 776 & 0.19 & 0.09 \\
	O & $\rightarrow$ & O & 5 & 0.21 & 0.10 \\
	S, SQ, SV, Sa, Sk,\\Sl, Sp, Sq, Sqw, Sr,\\Srw, Sv, Sw & $\rightarrow$ & S & 8,875 & 0.26 & 0.08 \\
	A, AQ & $\rightarrow$ & A & 69 & 0.28 & 0.09 \\
	Q, QO, QV & $\rightarrow$ & Q & 185 & 0.28 & 0.11 \\
	V, Vw & $\rightarrow$ & V & 1,412 & 0.36 & 0.11 \\
	E, X, XD, Xc, Xe, Xn, Xt & $\rightarrow$ & E & 46 & 0.46 & 0.16 \\
	X, XD, XL, Xe,\\Xk, Xt& $\rightarrow$ & X & 202 & - & - \\
	\bottomrule
\end{tabular}

%% file: tables/complexes.tex
\begin{tabular}{lrrrrrrrrrrr}	\toprule
	$\Sigma$& $N_c$ & $N_o$ & $C_c$ & $C_o$ & $A_c$ & $A_o$& $k_{c}$ & $k_{o}$ & $\zeta-1_{c}$ & $\zeta-1_{o}$ & $p$\\
	\midrule
	P & 255 & 576 & (0.80, 0.05) & (0.83, 0.06) & 4.0 & 6.4 & -1.82 & -1.81 & 0.16 & 0.12 & 0.00  \\
	D & 179 & 419 & (0.77, 0.17) & (0.72, 0.20) & 8.5 & 10.6 & -1.67 & -1.62 & 0.06 & 0.09 & 0.00  \\
	Ch & 125 & 255 & (0.77, 0.05) & (0.76, 0.07) & 4.1 & 5.2 & -1.84 & -1.80 & 0.22 & 0.21 & 0.02  \\
	B & 172 & 519 & (0.82, 0.06) & (0.77, 0.08) & 4.5 & 8.0 & -1.82 & -1.79 & 0.14 & 0.17 & 0.00  \\
	C & 965 & 3,609 & (0.82, 0.06) & (0.83, 0.06) & 6.2 & 5.0 & -1.81 & -1.82 & 0.13 & 0.13 & 0.00  \\
	T & 30 & 62 & (0.65, 0.19) & (0.53, 0.24) & 6.3 & 7.5 & -1.61 & -1.49 & 0.18 & 0.29 & 0.66  \\
	M & 203 & 642 & (0.19, 0.34) & (0.07, 0.42) & 9.0 & 7.5 & -1.05 & -0.77 & 0.92 & 1.02 & 0.00  \\
	K & 147 & 566 & (0.18, 0.40) & (0.06, 0.48) & 8.6 & 6.6 & -0.99 & -0.72 & 0.71 & 0.87 & 0.00  \\
	L & 176 & 758 & (0.16, 0.37) & (0.06, 0.47) & 9.0 & 6.7 & -0.96 & -0.73 & 0.89 & 0.89 & 0.00  \\
	S & 2,076 & 8,702 & (0.08, 0.46) & (0.04, 0.51) & 6.4 & 3.5 & -0.76 & -0.67 & 0.87 & 0.81 & 0.00  \\
	A & 17 & 68 & (0.30, 0.39) & (0.05, 0.57) & 7.5 & 6.2 & -1.16 & -0.68 & 0.46 & 0.60 & 0.22  \\
	Q & 14 & 184 & (0.36, 0.44) & (0.05, 0.52) & 9.2 & 4.6 & -1.18 & -0.70 & 0.25 & 0.74 & 0.02  \\
	V & 254 & 1,371 & (0.10, 0.56) & (0.04, 0.58) & 6.5 & 3.2 & -0.78 & -0.67 & 0.50 & 0.60 & 0.00  \\
	E & 19 & 43 & (0.33, 0.45) & (0.06, 0.48) & 8.0 & 8.8 & -1.14 & -0.73 & 0.29 & 0.86 & 0.23  \\
	X & 31 & 200 & (0.11, 0.45) & (0.06, 0.52) & 9.0 & 5.6 & -0.83 & -0.70 & 0.81 & 0.73 & 0.04  \\
	\bottomrule
\end{tabular}

%% file: sections/section5.tex
\section{Asteroid families in phase space}%
\label{sec:families}
In the following, we illustrate the use of the \GG phase curve coefficients as
an extension of the physical parameter space of families. We intend this as a
proof-of-concept of the results in \Autoref{sec:taxonomy} rather than a full
analysis of the implications.

The identification of asteroid families requires accurate parameter derivation
and large number statistics to discern their members from the background of
minor bodies \citep{10.1016j.icarus.2014.05.039}. It is an interplay of their
dynamical parameter space, specifically the proper orbital elements, and their
physical parameters such as albedos and colours
\citep{10.1086344077,10.1016j.icarus.2008.07.002,10.10880004-637x77017}

The phase curve coefficients represent a large corpus of physical quantities
when derived from serendipitous observations.
\citet{10.1016j.jqsrt.2011.03.003} compute family-specific phase curves by
fitting family members with constant \GG-parameters, minimizing a global
$\chi^2$ in a grid search and describing the quality of all fits simultaneously
to arrive at the best fit for the family collective. The resulting \GG-values
are concentrated towards medium photometric slopes and opposition effect sizes
for all 17 families in the study, among which is the high-albedo
\nuna{4}{Vesta}-family. We interpret this as indication that the simultaneous
treatment of all family members suppresses the inherent information on family
taxonomy and fraction of interlopers. 

\begin{figure*}[!th]
  \centering
  \input{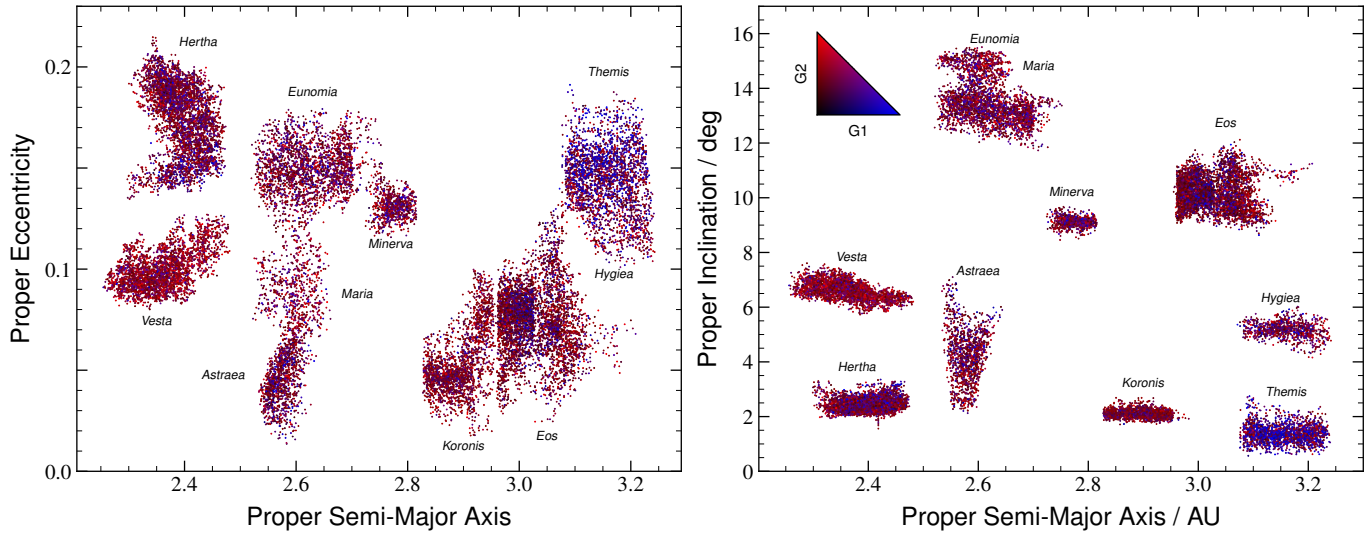}
  \caption{Illustrated are the \GG-parameters of several asteroid families,
    plotted in proper orbital elements space as semi-major-axis versus eccentricity
    (top) and versus inclination angle (bottom). The phase curves were observed by
    ATLAS in \orange. The proper orbital elements and family memberships are
    provided by \texttt{AstDyS-2} \citep{10.1016j.icarus.2014.05.039}. The
    colour-coding of the \GG-space for both figures is given in the inset of the
  right-hand plot.}
  \label{fig:g1g2_aei}
\end{figure*}

\begin{table*}[t]
  \centering
  \caption{Asteroid families with the number $N$ of members, the geometric
    centers C of the 95\% probability-level contour, the area $A$ of the
    1\,$\sigma$-contour, as observed by ATLAS in \cyan (subscript $c$) and \orange
    (subscript $o$). The areas $A$ are multiplied by 1,000 for notation purposes.
    Further given are the taxonomic classifications of the families and their
  references.}
  \input{tables/families.tex}
  \label{tab:families}
\end{table*}

The distribution of family members in \GG-space can yield insights on the nature
of their parent body or bodies. Unimodal distributions suggest a homogeneous
taxonomy, e.g., from a homogeneous single parent body or from compositionally
similar parent bodies of overlapping families.  A heterogeneous taxonomy in
either case would give rise to multimodal distributions in \GG-space, as would
the presence of a considerable fraction of interlopers. Finally, the
superposition of distinct families in orbital space could be reflected in their
phase coefficients in the case of different taxonomic nature.

We retrieve the proper orbital elements (semi-major axis $a_p$, eccentricity
$e_p$, and orbital inclination angle $i_p$, refer to
\citet{10.10160019-10359290091-k, 10.1023a:1011187405509}) for
\numbproperelements asteroids observed by ATLAS as well as their family
memberships from the \emph{Asteroids - Dynamic Site 2}
(\texttt{AstDyS-2})\,\footnote{\url{https://newton.spacedys.com/astdys2/}}
\citep{10.1016j.icarus.2014.05.039}. We select all families of which more than
\fammemberlimit members have been observed by ATLAS, either in \cyan or in
\orange, after applying the limit of $N\geq150$ on the sample from
\Autoref{sec:results}. \numbfamiliesoverlimit families pass the required number
of observed members: \listoffamilies.\,\footnote{Note that the \nuna{8}{Flora}
  family is not present in \texttt{AstDyS-2} as it does not differentiate
  sufficiently from the background in the hierarchical clustering method used, see
\citet{10.1016j.icarus.2014.05.039}.} The distributions of the families in
$a_p-e_p$- and $a_p-i_p$-space are shown in \Autoref{fig:g1g2_aei}. Each dot
represents an asteroid, colour-coded by its \GG-values. The illustrated sample
is restricted to phase curves observed in \orange to show the larger subsample
while eliminating the wavelength-dependency.  We quantify the \GG-distributions
for the families as done in \Autoref{sec:taxonomy} for the taxonomic complexes.
The 2\,D KDEs are depicted in \Autoref{fig:g1g2_taxonomy_families_all}, split
into \cyan and \orange. We summarize them in \Autoref{tab:families}, giving the
area of the 1\,$\sigma$-contour and the geometric center of the 95\%
probability-level contour. The former is indicative of the fraction of
interlopers or the taxonomic heterogeneity of the parent bodies, while the
latter characterizes the \GG-values of the core family members. In addition, we
state reference taxonomic classifications of the families.

Three families show strong uniformity, both visually in \Autoref{fig:g1g2_aei}
and in their small area sizes in \Autoref{tab:families}. \nuna{4}{Vesta} is the
archetype of the high-albedo taxonomic class, the V-types \citep{10.1086115658},
in agreement with the large $G_2$ values of its core member center positions.

\nuna{24}{Themis} is a C-type family with known low-albedo interlopers such as
the B-type subfamily \nuna{656}{Beagle}
\citep{10.1016j.icarus.2004.10.002,10.1016j.icarus.2016.01.002}. While in \cyan
these complexes appear indistinguishable, the blue B-types separate from the red
C-types towards the \orange wavelength-regime, refer to \Autoref{tab:complexes}.
We are not able to resolve this shift using the classified B- and C-types in the
\nuna{24}{Themis} family subsample, nevertheless, phase curves from targeted
observations might detect this difference \citep{10.1016j.pss.2015.11.007}.

\nuna{158}{Koronis} is one of the largest families in terms of number and we
confirm here its homogeneous S-type taxonomy \citep{1984PhDT.........3T}. 

The C-type family \nuna{10}{Hygiea} shows a considerable fraction of objects
with high albedos in \cyan, as well as objects with high albedos in \orange.
\citet{10.1093mnrasstt437} have identified S- and X-type interlopers in the
family. We further attribute this partially to remaining phase curves with
insufficient opposition effect coverage, as the distribution shifts towards
C-type objects with increasing $N$.

The C-type asteroid \nuna{93}{Minerva} is the namesake of an S-type family
\citep{10.1016j.icarus.2004.10.002}.

We note that the family centers given in \Autoref{tab:families} are not
compatible with the results of \citet{10.1016j.jqsrt.2011.03.003}; the geometric
centers of the families extend more towards the upper $G_1$- and $G_2$-values as
seen by \citet{10.1016j.pss.2015.08.010}.  We attribute this to our treatment of
each family member separately, allowing for a differentiated look into the
\GG-distributions, specifically separating the core family members and potential
interlopers.

As in \Autoref{subsec:identification_interlopers_gg}, we conclude here that the
\GG-space is well suited for interloper detection, adding a physical parameter
space to asteroid families that can confirm dynamical identification and
strengthen the definition of families. This, in turn, improves their age
estimates \citep{10.1016j.icarus.2015.04.041}.

%% file: tables/families.tex
\begin{tabular}{lrrrrrrrr}	\toprule
	Family & $N_c$ & $N_o$ & $C_c$ & $C_o$ & $A_c$ & $A_o$ & Class & Reference\\
	\midrule
	\nuna{4}{Vesta} & 229 & 1647 & (0.07, 0.50) & (0.04, 0.55) & 162 & 106 &  \taxVesta & \citet{\refVesta} \\
	\nuna{5}{Astraea} & 59 & 524 & (0.11, 0.48) & (0.07, 0.48) & 197 & 156 &  \taxAstraea & \citet{\refAstraea} \\
	\nuna{10}{Hygiea} & 101 & 473 & (0.75, 0.11) & (0.08, 0.44) & 191 & 184 &  \taxHygiea & \citet{\refHygiea} \\
	\nuna{15}{Eunomia} & 383 & 1647 & (0.11, 0.43) & (0.06, 0.49) & 183 & 170 &  \taxEunomia & \citet{\refEunomia} \\
	\nuna{24}{Themis} & 528 & 1218 & (0.80, 0.05) & (0.73, 0.08) & 96 & 151 &  \taxThemis & \citet{\refThemis} \\
	\nuna{93}{Minerva} & 114 & 539 & (0.08, 0.49) & (0.07, 0.49) & 159 & 170 &  \taxMinerva & \citet{\refMinerva} \\
	\nuna{135}{Hertha} & 264 & 1777 & (0.36, 0.34) & (0.07, 0.49) & 172 & 131 &  \taxHertha & \citet{\refHertha} \\
	\nuna{158}{Koronis} & 502 & 1333 & (0.06, 0.46) & (0.03, 0.52) & 122 & 71 &  \taxKoronis & \citet{\refKoronis} \\
	\nuna{170}{Maria} & 100 & 472 & (0.19, 0.41) & (0.05, 0.47) & 228 & 169 &  \taxMaria & \citet{\refMaria} \\
	\nuna{221}{Eos} & 697 & 2732 & (0.13, 0.36) & (0.04, 0.44) & 174 & 134 &  \taxEos & \citet{\refEos} \\
	\bottomrule
\end{tabular}

%% file: sections/section6.tex
\section{Error sources in serendipitous phase curves}
\label{sec:uncertainty_budget}

Reduced phase curves from targeted campaigns are available for in the order of
100 asteroids. To increase the number of asteroids with available phase
coefficients, exploiting serendipitous asteroid observations is necessary.  The
uncertainty $\sigma$ of the reduced magnitudes in serendipitous phase curves is
a propagation of uncertainties arising from their 3\,D-shape and from the
observational parameters themselves,
\begin{equation}%
  \label{equ:sources_uncertainty}
  \sigma \propto \sqrt{ \sigma_{\mathrm{PHOT}}^2 +
    \sigma_{\mathrm{PREC}}^2 +
    \sigma_{\mathrm{SYS}}^2 +
    \sigma_{\mathrm{ROT}}^2 +
    \sigma_{\mathrm{APP}}^2
  }\,,
\end{equation}
where $\sigma_{\mathrm{PHOT}}$ is the photometric uncertainty of a single
observation, $\sigma_{\mathrm{PREC}}$ refers to the loss in precision when
magnitudes are reported in a truncated format, $\sigma_{\mathrm{SYS}}$ is
introduced by varying photometric systems used either in different observatories
or by an observatory over time, and $\sigma_{\mathrm{ROT}}$ and
$\sigma_{\mathrm{APP}}$ are magnitude modulations introduced by the asteroid's
shape, specifically the asteroid's rotation and the change in aspect angle over
different apparitions respectively. These uncertainties, disperse the observed
reduced magnitudes, leading to broader posterior distributions
(i.e.\,uncertainties) of the phase curve parameters as seen in
\Autoref{fig:eichsfeldia_phase_fit}. \Autoref{equ:sources_uncertainty} is a
non-exhaustive list, though the dominating error sources are encompassed.

Following the discussions in \Autoref{sec:results} and \Autoref{sec:taxonomy},
the coverage of the opposition effect further affects the derived
\HGG-parameters. As opposed to the factors in \Autoref{equ:sources_uncertainty},
its impact can be minimized by a carefully set observation schedule.

In the following, we examine the order of magnitude of each uncertainty listed
in \Autoref{equ:sources_uncertainty} and different degrees of coverage of the
opposition effect, intending to identify the dominating uncertainty and means to
minimize it. 

\begin{figure}[t]
  \centering
  \input{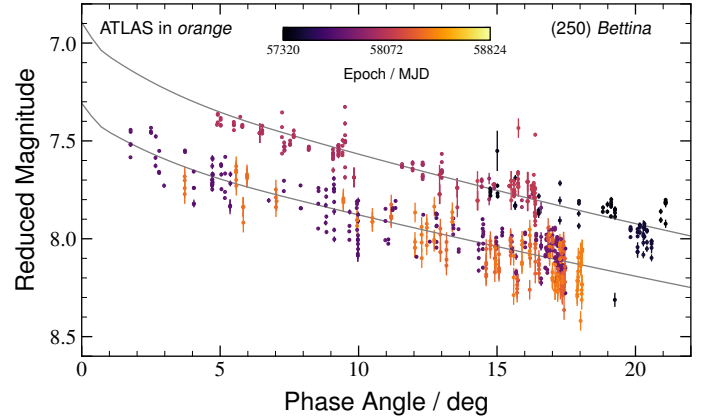}
  \caption{The phase curve of \nuna{250}{Bettina} as observed by ATLAS in
    \orange. The observations are colour-coded by their epoch, highlighting the four
    different apparitions that were captured. The triaxial ellipsoid ratios of
    \nuna{250}{Bettina} are 1.4:1:1 \citep{10.10510004-6361201731456}. The gray
  lines show the \HGG-model fits to the apparitions, split into pairs of two.}%
  \label{fig:bettina_apparitions}
\end{figure}

\subsection{Photometric Uncertainty and Precision}%
\label{sub:precision}

In the most basic form, any observed magnitude carries an uncertainty e.g.\,due
to random photon noise. These cannot be avoided and are present also in targeted
campaigns. In the ATLAS observations, the mean photometric error is 0.14\,mag,
with a standard deviation of 0.08\,mag depending largely on the apparent
magnitude of the target.  The LSST aims at $\sigma_{\mathrm{PHOT}}\sim0.01$\,mag
for single exposures of objects with a magnitude in $r$ of 21
\citep{2009arXiv0912.0201L}. 

When working with serendipitous observations, however, this precision is
frequently truncated to 0.1\,mag either when the observations are reported to or
retrieved from the MPC\@. This adds an uncertainty of
$\sigma_{\mathrm{PREC}}=0.1/\sqrt{12}$\,mag on top of $\sigma_{\mathrm{PHOT}}$,
where the $\sqrt{12}$ divisor comes from the standard deviation of the uniform
distribution. Once the new data pipeline accepting the updated observation
report format has been put into place by the MPC, this source of error will be
removed \citep{2017DPS....4911214C}.

\subsection{Photometric systems}%
\label{sub:observation_uncertainty}

There are systematic magnitude offsets which have to be taken into account when
combining magnitudes observed by different observatories, or even data from a
single observatory which underwent recalibration. A more in-depth look is done
by \citet{10.1016j.jqsrt.2011.03.003}. The differences in the photometric
systems are not always apparent, e.g.\,when retrieving observations from the
MPC.

As a baseline estimation of $\sigma_{\mathrm{SYS}}$, we use the example of the
SDSS and Pan-STARRS systems. Both share the $g$, $r$, $i$, and $z$ filters,
though there are slight differences in their throughputs. Computing the average
differences in apparent magnitude of the mean spectra for the 24 taxonomic
classes in \citep{10.1016j.icarus.2009.02.005} gives $\sigma_\mathrm{SYS}$
values of 0.09\,mag, 0.01\,mag, 0.02\,mag, and 0.08\,mag for the four filters
respectively. For bright asteroids, this is in the order of
$\sigma_{\mathrm{PHOT}}$.

\begin{figure}[t]
  \centering
  \input{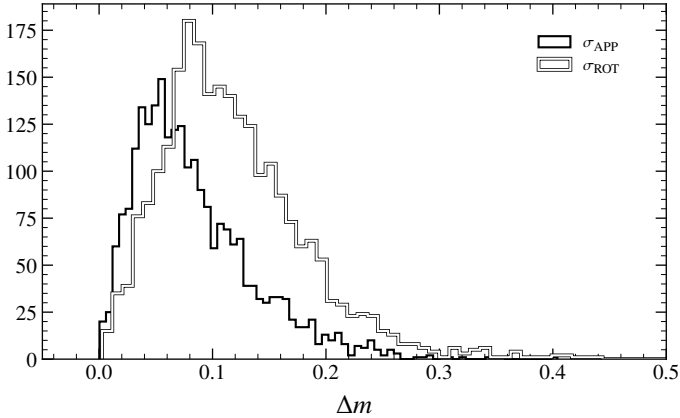}
  \caption{The amplitude of magnitude dispersion due to the asteroids' rotation
    (black) and change in aspect angle between different apparitions (white). The
    values are derived using the DAMIT shape models from
  \citet{2010A&A...513A..46D}.}%
  \label{fig:smag_dmag}
\end{figure}

In \Autoref{sub:ioe}, we highlight the importance of densely sample phase
curves. We see here that achieving a large number of observations by combining
data from different photometric systems is a trade-off between increased phase
curve coverage and introduced dispersion in apparent magnitude.

\subsection{Asteroid rotation and apparition effect}%
\label{sub:rotation_and_apparition}

The light curve of an asteroid is modulated by to its 3\,D-shape rotating around
its spin axis. The rotation imprints a periodic modulation of the apparent
magnitude over the rotation period, which is typically in the order of a few
hours \citep{10.1016j.icarus.2009.02.003}.\,\footnote{\url{http://alcdef.org/}}
In addition, the varying aspect angles over different apparitions of the
asteroid introduce offsets in the observed magnitude, effectively shifting the
whole phase curve along the $y$-axis, hence biasing the determination of the
absolute magnitude $H$.

\Autoref{fig:bettina_apparitions} shows an example of both effects affecting the
phase curve of \nuna{250}{Bettina}, as observed by ATLAS in \orange. The epoch
of observation is colour-coded. Four distinct apparitions can be seen in the
observations, leading to shifts in the reduced magnitudes which give the
impression of two superimposed phase curves being displayed. On top of the
apparition effect, we see the magnitude dispersion within the apparitions,
introduced in part by the asteroid's spin. 

The strength of the magnitude modulation due to rotation and change in aspect
angle depend on the shape of the asteroids and the viewing geometry. The
\emph{Database of Asteroid Models from Inversion Techniques}
(DAMIT)\,\footnote{\url{http://astro.troja.mff.cuni.cz/projects/damit}} provides
shape models for \numbdamitasteroids asteroids \citep{2010A&A...513A..46D}. We
use these shape models to quantify the effect of the rotational modulation
($\sigma_{\mathrm{ROT}}$) and of the varying aspect angle
($\sigma_{\mathrm{APP}}$).

For each asteroid, we compute its triaxial dimensions ($a>b>c$) and assimilate
its shape to a smooth ellipsoid in the following. Under this assumption, the
modulation of the apparent magnitude due to spin and 3D shape writes
\begin{equation}
  \begin{split}
    m &= -2.5 \log\Bigg(  \pi a b c \cdot\\
      &\Bigg(\left(\frac{\cos{\beta}\cos{\lambda}}{b}\right)^2 + \left(
          \frac{\cos{\beta}\sin{\lambda}}{a}\right)^2 + \left(
    \frac{\sin{\beta}}{c}\right)^2\Bigg)^{0.5}\Bigg)\,,
    \end{split}
  \end{equation} 
  with $\lambda$, $\beta$ being the longitude and latitude of the subobserver
point \citep{1978A&A....66...31S,10.10160019-10358490129-5}.

We generate a full-rotation synthetic light curve every 10 days over an entire
orbital revolution around the Sun, effectively probing the range of
Sun-target-observer geometries. As serendipitous observations randomly occur
over the rotation period, the measured magnitude is offset from the average
value. We estimate this offset $\sigma_{\mathrm{ROT}}$ by computing the root
mean square residuals of each light curve to its average.

The influence of the varying aspect angle is computed from the difference
$\sigma_{\mathrm{APP}}$ between the average magnitude of each light curve and a
light curve taken while the observer is located within the equatorial plane of
the asteroid.

The distributions of the changes in apparent magnitude for both effects are
shown in \Autoref{fig:smag_dmag}. $\sigma_\mathrm{ROT}$ is in general larger
than $\sigma_\mathrm{APP}$, with a median value of 0.11\,mag compared to
0.07\,mag respectively. For $\sigma_\mathrm{APP}$, the situation is analogous to
$\sigma_\mathrm{SYS}$; it can be removed by avoiding the combination of
observations from different apparitions of the target. Nevertheless, the
benefits of adding more samples of the phase curve may outweigh the downsides of
increased magnitude dispersion. In \Autoref{fig:bettina_apparitions}, the
parameter inference failed when applied to all observations, while we could
retrieve two phase curves of \nuna{250}{Bettina} after splitting the apparitions
in pairs of two.

\subsection{Effect of magnitude dispersion on \HGG}%
\label{sub:quantifying_the_influence_of_dispersion}

To quantify how the magnitude dispersion due to the uncertainties listed in
\Autoref{equ:sources_uncertainty} affect the \HGG-model parameters, we simulate
the \nuna{20}{Massalia} phase curve by \citet{10.1086146166} in
\Autoref{fig:massalia_phase_fit} as serendipitous observations. Using the order
of magnitudes for the uncertainties derived above
($\sigma_\mathrm{PHOT}=0.1$\,mag, $\sigma_\mathrm{PREC}=0.1/\sqrt{12}$\,mag,
$\sigma_\mathrm{SYS}=0.05$\,mag, $\sigma_\mathrm{ROT}=0.11$\,mag, and
$\sigma_\mathrm{APP}=0.07$\,mag), we compute the propagated uncertainty and
simulate 100 phase curves of \nuna{20}{Massalia} with $N=6$ at the same phase
angles as the original observations, but randomly displaced following a Gaussian
distribution with the mean value of the original magnitude observed at the
respective phase angle and the standard deviation of the propagated uncertainty
$\sigma$.
For these 100 phase curves, we compute the \HGG-parameters and the differences
to the parameters of the original phase curve,
\begin{equation}%
  \label{equ:deltas}
  \Delta\boldsymbol{\Theta} = \boldsymbol{\Theta}_i -
  \boldsymbol{\Theta}\mathrm{,\quad i\,\in\,\{1, \dots, 100\}}\,,
\end{equation}
where $\boldsymbol{\Theta}$ refers to the \HGG-parameters.  The mean difference
and the median value of the absolute difference of the resulting distributions
are given in the first row of \Autoref{tab:deltas}. The former indicates
systematic parameter shifts with respect to the original values, while the
latter indicates the spread of the differences. We choose the median rather than
the standard deviation to be able to compare the results to the non-Gaussian
distributions we obtain in the next subsection.

\begin{table}[t]
  \centering
  \caption{The distributions of the differences between the 100 phase curves
    with simulated noise ($\sigma$) and the \HGG-parameters of the targeted
    \nuna{20}{Massalia} observations by \citet{10.1086146166}. The same is given for
    the truncated ATLAS phase curves, with $i$\,deg describing the dropout degree.
    $\mu$ refers to the mean values of the distributions, while $\sigma$ gives the
  median of the absolute differences.}
  \input{tables/deltas.tex}
  \label{tab:deltas}
\end{table}

The simulated phase curves on average show larger $H$ and $G_1$, i.e.\,they are
steeper while depicting smaller opposition effect sizes.  While the opposition
effect in the original observations is sampled sufficiently, the stochastic
nature of the simulated observations is reflected in the large median values of
the absolute differences of all three parameters.  This highlights the need for
dense sampling to allow for restricting the parameter space; the offset in
$\mu\Delta H$ in the simulated phase curves renders taxonomic classification
from the computed colours inconclusive.

\subsection{Opposition effect coverage}%
\label{subsec:dependency_on_phase_angle_coverage}

When relying on serendipitous asteroid observations, the coverage of the
opposition effect is pre-determined by the survey footprint. This dependence of
the scientific yield in terms of asteroid phase curves, colours, or taxonomies
on the solar elongation coverage should be taken into account early on.

To quantify the influence of insufficient phase angle coverage, we select all
ATLAS phase curves in \cyan and \orange with $\alpha_\mathrm{min}\leq1$\,deg and
$N\geq$~50 of asteroids with shape models present in DAMIT.
\numbdamitatlasphasecurves phase curves of \numbdamitatlas asteroids fulfill
these criteria.  Next, we reduce the spin- and apparition-induced magnitude
dispersion in the phase curves using shape models and the light curve generation
software\,\footnote{\url{https://astro.troja.mff.cuni.cz/projects/damit/pages/software_download}}
provided by DAMIT. This increases the probability that in the following
simulations, we observe the influence of the opposition effect coverage rather
than the magnitude dispersion on the \HGG-parameters. However, the photometric
noise cannot be removed by essence and some residuals arise from the non-ideal
fit of the photometry by the shape-induced light curve.

\begin{figure}[t]
  \centering
  \input{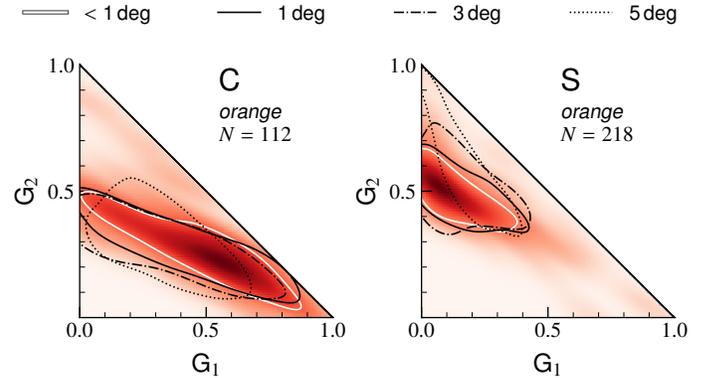}
  \caption{The effect of insufficient phase curve coverage at low angles on
    the \GG-parameters. Shown are the 2\,D fitted kernel density estimators of the
    \GG-distribution for C-type and S-type asteroids (red). The solid white line
    displays the 1\,$\sigma$-level of the distribution using the complete ATLAS
    phase curves, the black solid, dash-dotted, and dotted contours the
    distributions using the phase curves truncated at 1\,deg, 3\,deg, and 5\,deg
  respectively.}
  \label{fig:g1g2dropout}
\end{figure}

The mean minimum phase angle of the reduced subset is 0.6\,deg. We remove
observations below \{1, 2, 3, 4, 5\}\,deg phase angle and compute the \HGG-model
fit. The relevance of the opposition effect can then be quantified by looking at
the difference of the \HGG-parameters of the complete and the truncated phase
curves. In \Autoref{tab:deltas}, we display the difference in the parameters of
the truncated phase curves and the complete phase curves, defined analogously to
\Autoref{equ:deltas} with $i$ referring to the truncation angle. With
diminishing coverage of the opposition effect, the error on $H$ increases up to
0.2\,mag in these simulations. Following \autoref{equ:hdp}, this translates to
an error of 17\% on the derived asteroid albedo. Further, we observe systematic
shifts of the \GG parameters. We display this in \Autoref{fig:g1g2dropout},
depicting the \GG-distributions for 112 C-types and 218 S-types observed in
\orange. The contours depict the 1\,$\sigma$-outlines for the complete phase
curves (white), and increasing truncation angle (black). As the angle increases,
the taxonomic information gets lost.

%% file: tables/deltas.tex
\begin{tabular}{lrrrrrr}	\toprule
	 & $\mu_{\Delta_H}$ & $\sigma_{\Delta_H}$ &    $\mu_{\Delta_{G_1}}$ & $\sigma_{\Delta_{G_1}}$ & $\mu_{\Delta_{G_2}}$ & $\sigma_{\Delta_{G_2}}$ \\
	\midrule
	 $\sigma$ & 0.04 & 0.11 & 0.07 & 0.08 &  -0.01 & 0.10\\
\rule{0pt}{0.05cm}\\
	 1\,deg & 0.00 & 0.07 & -0.00 & 0.04 &  0.01 & 0.01\\
	 2\,deg & -0.00 & 0.12 & -0.01 & 0.07 &  0.01 & 0.02\\
	 3\,deg & -0.01 & 0.15 & -0.02 & 0.09 &  0.02 & 0.03\\
	 4\,deg & -0.00 & 0.18 & -0.03 & 0.09 &  0.03 & 0.03\\
	 5\,deg & 0.02 & 0.21 & -0.03 & 0.10 &  0.05 & 0.05\\
	\bottomrule
\end{tabular}

%% file: sections/section7.tex
\section{Conclusion}%
\label{sec:conclusion}

We perform phase curve parameter inference using serendipitous asteroid
observations for a large number of minor bodies. The ATLAS observatory provided
us with dual-band photometry for more than 180,000 objects, of which we selected
about 95,000 based on the sampling statistics of their phase curves. As ATLAS
continues to survey the night sky, we will be able to add an increasing number
of asteroids to this analysis.

Our results show that the \HGG-model parameters contain significant taxonomic
information of the target surface, provided the opposition effect is densely
sampled. The close correlation between the \GG-parameters and the albedo allows
to use serendipitously observed phase curves as accessible albedo proxy for
hundreds of thousands of asteroids with the upcoming LSST and NEOSM surveys.
The taxonomic complexes separate sufficiently in the phase-coefficient space to
study ensembles of asteroids such as asteroid families, while the large tails of
the distributions prevent classification of single objects from \GG alone.

We find evidence for a wavelength-dependency of the phase coefficients.
Provided the taxonomy is known, the derived slope parameters of the complexes
allow for estimating the degree of phase reddening in the slopes of asteroid
spectra.

We quantified the sources of uncertainties of serendipitously acquired phase
curves and their effect on the \GG parameters.  By simulating incomplete phase
curves at low phase angles, we highlight the importance of observations close to
opposition ($\leq$ 1\,deg) to determine the fundamental absolute
magnitude $H$, used to derived properties such as albedo, colours, and taxonomic
class.

%% file: frontback/ack.tex
\section*{Acknowledgements}%
\label{sec:acknowledgements}

This work has made use of data from the Asteroid Terrestrial-impact Last Alert
System (ATLAS) project. ATLAS is primarily funded to search for near earth
asteroids through NASA grants NN12AR55G, 80NSSC18K0284, and 80NSSC18K1575;
byproducts of the NEO search include images and catalogs from the survey area.
The ATLAS science products have been made possible through the contributions of
the University of Hawaii Institute for Astronomy, the Queen’s University
Belfast, the Space Telescope Science Institute, and the South African
Astronomical Observatory.

% Miriade
This research has made use of IMCCE's Miriade VO tool.

% SVO FILTER
This research has made use of the SVO Filter Profile Service
supported from the Spanish MINECO through grant AYA2017-84089.

%% file: frontback/appendix.tex
% Proper referencing of figures and tables
\renewcommand\thefigure{\thesection.\arabic{figure}}
\renewcommand\thetable{\thesection.\arabic{table}} 
\setcounter{figure}{0}
\setcounter{table}{0}

\appendix
\onecolumn

\section{Online catalogue}
\label{app:online_catalgoue}

We briefly describe the structure of the online catalogue submitted to the CDS
of phase curve parameters. The results are provided in a CSV-formatted table
consisting of \numbphasecurves rows, with each row providing the parameters of a
single asteroid in a single ATLAS observation band. About a third of the
asteroids appear in twice, once for \cyan and once for \orange. The column names
and brief descriptions are given in \Autoref{tab:online_catalogue}.  For
\numbfailedfits \HGG-model fits, no MCMC sample was withing the physical range
of the \GG-parameters. These fits are considered failed and the parameters are
empty in the catalogue provided online.

\begin{table}[h]
  \centering
  \caption{Structure of the online catalogue providing the phase curve parameters.}
  \label{tab:online_catalogue}
  \input{tables/catalogue}
\end{table}

\section{Observation bias towards large phase angles for impactor detection}
\begin{figure}[h]
  \centering
  \input{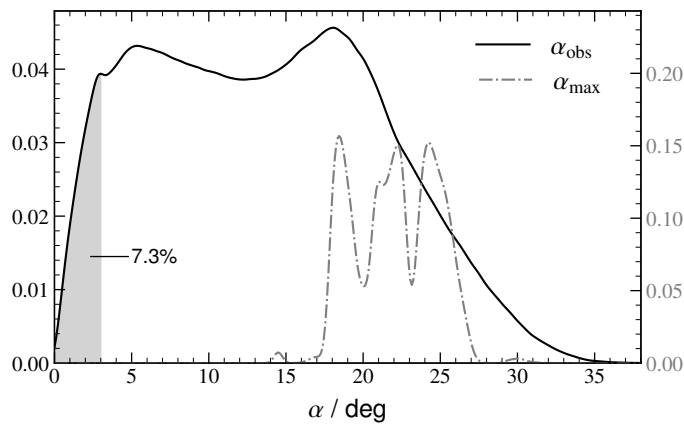}
  \caption{The black line shows the kernel density estimation of the
    distribution of asteroid phase angles at the epoch of observations, for all 20.7
    million ATLAS observations analysed in this work. We further show the
    kernel density estimation of the distribution of the maximum observable phase
    angles of all asteroids in the sample, derived from their proper semi-major axis
  $a_p$ as $\alpha_\mathrm{max} = 2/\sin^{-1}{(1/(2a_p))}$.}
  \label{fig:phase_angle_dist}
\end{figure}

\clearpage

\section{\GS of taxonomic complexes}%
\label{sec:gs_of_taxonomic_complexes}
\begin{figure}[h]
  \centering
  \input{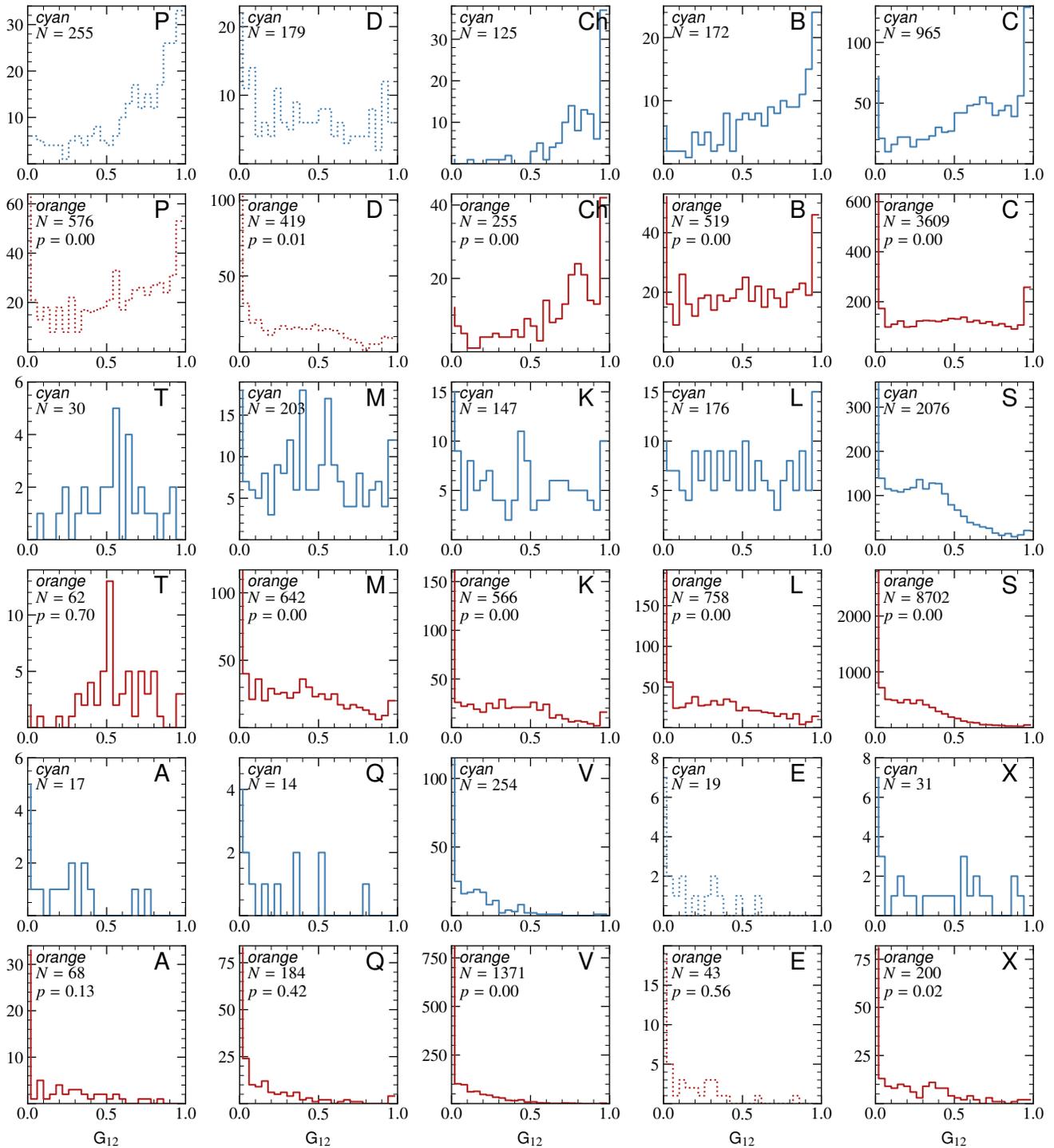}
  \caption{The \GS-distributions for several taxonomic complexes of
    asteroids, derived from serendipitous observations by ATLAS in \cyan and in
    \orange. We give the sample sizes $N$ and the two-sample Kolmogorov-Smirnov
    p-value between the \cyan and \orange \GS-distributions of the complexes. The
    \HGS-model is not suited for the D-, E-, and P-complexes, hence they are
  displayed with a dotted linestyle.}
  \label{fig:g12all}
\end{figure}

\clearpage

\section{\GG of families}%
\label{sec:gg_of_families}

\begin{figure}[h]
  \centering
  \input{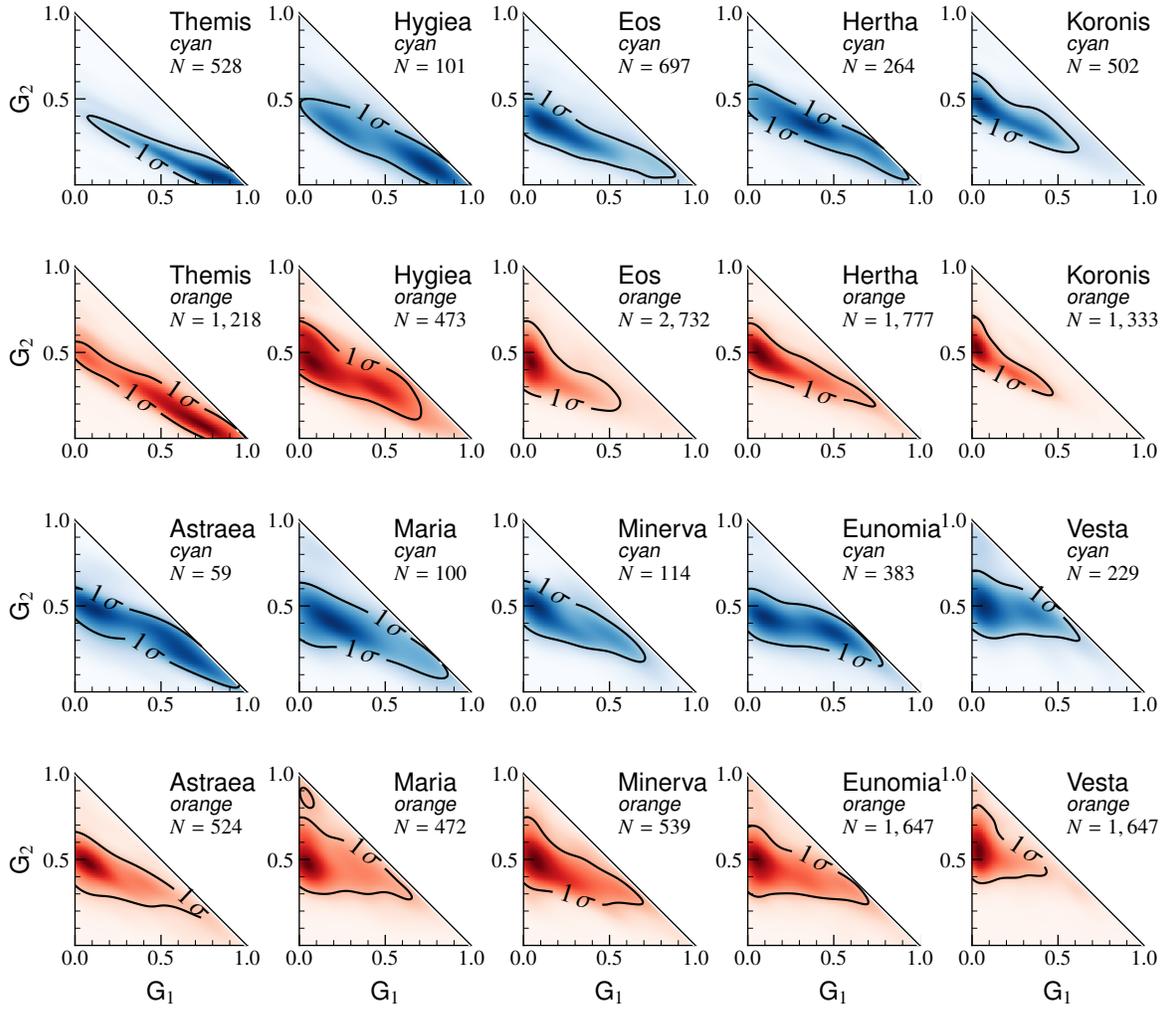}
  \caption{Depicted are the \GG-distributions of the members of 10
    asteroid families observed by ATLAS in \cyan (blue) and \orange (red). The black
    contours illustrate the 1\,$\sigma$-levels of the fitted 2\,D kernel density
    estimators.  The number of phase curves in each family and band is given by
  $N$.}
  \label{fig:g1g2_taxonomy_families_all}
\end{figure}

%% file: tables/catalogue.tex
\begin{tabular}{ll|ll}
  \toprule
  \texttt{Column}           & Description                                     &
  \texttt{Column}           & Description                                       \\
  \midrule
  \texttt{number}           & Asteroid number                                 &
  \texttt{rms12}            & RMS of
  \HGS-model fit                                                                \\

  \texttt{name}             & Asteroid name or designation                    &
  \texttt{h$\_$up}          & Upper 95\% HDI value of $H$                       \\

  \texttt{band}             & ATLAS observation band, either \cyan or \orange &
  \texttt{h$\_$low}         & Lower 95\% HDI value of $H$                       \\

  \texttt{class}            & Reference taxonomic classification for          &
  %\asteroidswithclass asteroids & 
  \texttt{g1$\_$up}         & Upper 95\% HDI value of $G_1$                     \\

  \texttt{scheme}           & Taxonomic scheme of reference                   &
  \texttt{g1$\_$low}        & Lower 95\% HDI value of $G_1$                     \\

  \texttt{ref$\_$tax}       & Code to identify taxonomy reference             &
  \texttt{g2$\_$up}         & Upper 95\% HDI value of $G_2$                     \\

  \texttt{ap}               & Proper semi-major-axis from
  \texttt{AstDyS-2}         &
  %for \numbproperelements asteroids  &
  \texttt{g2$\_$low}        & Lower 95\% HDI value of $G_2$                     \\

  \texttt{ep}               & Proper eccentricity from
  \texttt{AstDyS-2}         &
  %for \numbproperelements asteroids  &
  \texttt{h12$\_$up}        & Upper 95\% HDI value of $H_{12}$                  \\

  \texttt{ip}               & Proper inclination from
  \texttt{AstDyS-2}         &
  %for \numbproperelements asteroids  &
  \texttt{h12$\_$low}       & Lower 95\% HDI value of $H_{12}$                  \\

  \texttt{N}                & Total number of observations                    &
  \texttt{g12$\_$up}        & Upper 95\% HDI value of \GS                       \\

  \texttt{phmin}            & Minimum phase angle of observations             &
  \texttt{g12$\_$low}       & Lower 95\% HDI value of \GS                       \\

  \texttt{phmax}            & Maximum phase angle of observations             &
  \texttt{albedo}           & Reference albedo                                  \\
  %for \asteroidswithalbedo asteroids \\

  \texttt{h}                & Fitted $H$ of \HGG-model                        &
  \texttt{err$\_$albedo}    & Uncertainty of reference albedo                   \\

  \texttt{g1}               & Fitted $G_1$ of \HGG-model                      &
  \texttt{ref$\_$albedo}    & Code to identify albedo reference                 \\

  \texttt{g2}               & Fitted $G_2$ of \HGG-model                      &
  \texttt{family$\_$number} & Family number from
  \texttt{AstDyS-2}                                                             \\
  %for \asteroidswithfamilies asteroids  \\

  \texttt{rms}              & RMS of \HGG-model fit                           &
  \texttt{family$\_$name}   & Family name from
  \texttt{AstDyS-2}                                                             \\
  %for \asteroidswithfamilies asteroids \\

  \texttt{h12}              & Fitted $H_{12}$ of \HGS-model                   &
  \texttt{family$\_$status} & Family status from
  \texttt{AstDyS-2}                                                             \\
  %for \asteroidswithfamilies asteroids \\

  \texttt{g12}              & Fitted \GS of \HGS-model                        & \\
  \bottomrule
\end{tabular}